\documentclass[aps,pra,reprint, amsmath,amssymb,nofootinbib,longbibliography]{revtex4-2}
\usepackage{graphicx}
\usepackage{dcolumn}
\usepackage{bm}

\usepackage[T1]{fontenc}

\usepackage{braket}
\usepackage{dsfont}
\usepackage{booktabs}
\usepackage{placeins}

\usepackage{subcaption}
\usepackage[ruled,vlined]{algorithm2e}

\newcommand{\sbline}{\\[.5\normalbaselineskip]}

\newcommand{\plotsize}{0.25}

\begin{document}
	
\title{Deep reinforcement learning for key distribution based on quantum repeaters}
\author{Simon D. Reiß}
\email{sreiss@uni-mainz.de}
\author{Peter van Loock}
\email{loock@uni-mainz.de}
\affiliation{Johannes-Gutenberg University of Mainz, Institute of Physics, Staudingerweg 7, 55128 Mainz, Germany}

\begin{abstract}
This work examines secret key rates of key distribution based on quantum repeaters in a broad parameter space of the communication distance and coherence time of the quantum memories. As the first step in this task, a Markov decision process modeling the distribution of entangled quantum states via quantum repeaters is developed. Based on this model, a simulation is implemented, which is employed to determine secret key rates under naively controlled, limited memory storage times for a wide range of parameters. The complexity of the quantum state evolution in a multiple-segment quantum repeater chain motivates the use of deep reinforcement learning to search for optimal solutions for the memory storage time limits - the so-called memory cut-offs. The novel contribution in this work is to explore very general cut-off strategies which dynamically adapt to the state of the quantum repeater. An implementation of this approach is presented, with a particular focus on four-segment quantum repeaters, achieving proof of concept of its validity by finding exemplary solutions that outperform the naive strategies.
\end{abstract}
 
\maketitle

\section{Introduction}
\label{sec:introduction}
In the last century, the discovery and progress of quantum physics fundamentally reshaped the understanding of the world. In the 21st century, the scientific community started to utilize quantum mechanics to develop new technologies emerging as an entire field of quantum technology. In this field, quantum cryptography is of high relevance and the first technology to be on its way to broader commercialization. Common classical cryptography relies upon computational hardness assumptions to ensure the security of the transmitted information. This is an inherently vulnerable concept, since advances in the computational power can break the condition on which the security was built. Quantum cryptography, however, offers unconditional security based on fundamental laws of physics \cite{PhysRevLett.85.441,BENNETT20147}. In recent years, substantial research has been put into realizing practical quantum cryptography. While fiber-based quantum key distribution (QKD) over a couple of hundreds of kilometers has already been successfully implemented and employed \cite{PhysRevLett.117.190501,PhysRevLett.121.190502}, achieving high transmission rates over longer distances, up to 1000 km and beyond, remains an ambition for the foreseeable future. Complementary, satellite-based quantum communication transmitting quantum information through free space is another promising approach \cite{doi:10.1126/science.aan3211,PhysRevLett.115.040502}. Ultimately, global quantum networks are expected to be based upon a combination of fiber and satellite links. Another interesting option is the concept of twin-field QKD reaching distances beyond 400 km with standard optical fibers \cite{Lucamarini2018} and up to about 800 km by using low-loss fibers (0.1419 dB/km) as was recently demonstrated in Ref. \cite{Wang2022}.

The fundamental building block to extend the range of quantum networks to larger distances is a quantum repeater \cite{Muralidharan2016,PhysRevLett.81.5932}. Similar to classical communication, the optical fibers used in quantum communication suffer from channel losses that exponentially grow with distance. Quantum repeaters are designed to overcome these losses and preserve transmission rates at long distances by dissecting the communication channel, distributing entangled quantum states over sufficiently short segments, and eventually connecting the elementary links via quantum teleportation (entanglement swapping) \cite{PhysRevLett.81.5932}. The ongoing development of quantum repeaters has seen significant progress in recent years in the theoretical concepts \cite{Muralidharan2016,doi:10.1063/1.5115814,PhysRevResearch.1.023032,Coopmans2021,https://doi.org/10.48550/arxiv.2203.10318}, proposals for implementations \cite{Duan2001,PhysRevLett.96.240501,PhysRevLett.96.070504}, as well as the engineering of the necessary hardware \cite{https://doi.org/10.1002/qute.201900141,Humphreys2018,Bhaskar2020,PhysRevLett.126.230506}.

Besides the experimental implementations, suitable strategies operating the quantum key distribution over quantum repeaters have to be developed. This turns out to be a very challenging task, since the complexity of analyzing multiple-segment quantum repeaters grows quickly with the number of repeater stations and hence the distance \cite{PhysRevLett.128.150502,https://doi.org/10.48550/arxiv.2203.10318}.

Memory-based quantum repeaters store intermediate states in quantum memories and are currently the most experimentally feasible approach that is scalable to large distances by concatenating sufficiently many quantum repeater nodes \cite{https://doi.org/10.1002/qute.201900141,RevModPhys.83.33}. Hence, one task of the operating protocol is to manage and control the quantum states stored in the quantum memories. Physical implementations of today's quantum memories suffer from degradation of the stored quantum states \cite{Rozp_dek_2018,PhysRevA.99.052330,https://doi.org/10.1002/qute.201900141}.

There are at least two established, similar approaches to counteract the memory degradation and improve the fidelity of the distributed states at the cost of lower rates. One approach is based upon a so-called cut-off where quantum states are discarded when their storage times exceed a chosen threshold \cite{Rozp_dek_2018,PhysRevA.99.052330,PhysRevLett.98.060502,PhysRevA.100.032322,praxmeyer2013reposition,PhysRevResearch.1.023032,8972391,khatri2020policies,9495278}. An approach to simplify the computation is to use a memory buffer instead (sometimes referred to as "memory access time") where the generation of initial entangled states between adjacent repeater stations is restarted at fixed times \cite{Santra_2019,PhysRevA.103.032610}. A memory buffer is distinct from a memory cut-off, since two neighboring segments, even when ready, must wait until a predetermined time resulting in an unnecessary dephasing of the states. In the case of a cut-off, states that have waited for any duration below the cut-off are swapped as soon as possible.

The motivation of this work is to optimize the quantum repeater strategies in quantum key distribution tasks. Exact analytical or even numerical optimization approaches often seem computationally infeasible in the treatment of multiple-segment quantum repeaters \cite{8972391,PhysRevA.100.032322,Santra_2019,PhysRevA.99.042313,PhysRevA.103.032610,Jiang17291}. In the following, we will use the typical terminology of machine learning where strategies are termed policies.

In Ref. \cite{9495278} a numerical optimization of the cut-off to maximize the secret key rate is presented. In that work the cut-off is optimized per nesting level. The results presented in Ref. \cite{9495278} use a fixed ("doubling") swapping scheme which differs from the present work where quantum states are "swapped as soon as possible". More importantly, the policies presented in our work are more versatile in their ability to dynamically adapt to the state of the repeater chain and hence are significantly more complex to analyze.

On the way towards optimizing large-scale quantum networks incorporating a vast parameter space, methods able to handle this level of complexity remain of particular interest. Reinforcement learning (RL) is a method capable of finding near optimal solutions to problems where an analytical treatment is infeasible. Deep reinforcement learning (DRL) extends RL by the use of artificial neural networks. In recent years, DRL has made significant advances in optimizing control tasks for problems which where previously unsolvable \cite{8103164,duan2016benchmarking}. Notable examples include training a computer to play Atari games from raw game pixels \cite{Mnih2015} and performing locomotion tasks \cite{schulman2015trust,heess2015learning}. This motivates the application of these methods to quantum communication networks, offering solutions even for scenarios where the complexity exceeds what is achievable with other numerical optimization approaches.

Most recently, RL was successfully applied to some quantum information tasks. For example, in Refs. \cite{PRXQuantum.1.010301,PhysRevX.8.031084} agents autonomously developed well-known quantum information protocols and completed quantum error correction strategies, respectively. In Ref. \cite{Nautrup2019optimizingquantum} RL was used to optimize quantum error correction codes.

The above-mentioned references as well as the present work make use of classical algorithms to solve quantum problems. This should not be confused with quantum machine learning approaches where the optimization itself has quantum aspects to it.

In the present work, DRL will be applied to QKD via quantum repeaters \cite{masterReiss}. As the first step, we formulate a memory-based multi-segment quantum repeater as a Markov decision process (MDP). This MDP incorporates the full description of the quantum states and incorporates channel loss and Pauli errors as well as the option to discard any intermediate quantum state. Based on this, a simulation is employed, including a simple uniform memory cut-off. We present a broad range of results on the dependency of the secret key rate of the experimental parameters for the memories, the segment lengths, and the uniform cut-off parameter for the special class of four-segment quantum repeaters. The simulation also serves as the necessary groundwork for our DRL approach.

We adapt a public implementation of a proximal policy optimization (PPO) DRL algorithm to the simulation in order to find sophisticated memory policies optimizing the secret key rate. The major obstacle in this application of DRL is that the optimization merit is non-additive in terms of the fidelity of the distributed quantum states. We offer an elegant solution in proposing a generalized objective function, expanding common RL theory while maintaining the applicability of convergence improving techniques related to value functions. This improves computational feasibility compared to a simple solution via an episodic reward. The search space consists of the full memory control over discarding individual quantum states, based on the entire information available at any moment.

In this dynamic adapting of the policy lies the novel contribution of our work, compared to prior work which considered static fixed cut-offs that were assigned to nesting levels of a doubling scheme or a single point-to-point link. Furthermore, DRL is the enabling method to achieve this versatility. It has proven to excel in optimization tasks whose complexity exceeds the capabilities of other numerical approaches \cite{8103164,duan2016benchmarking}, thus offering an approach for large-scale networks including numerous interleaving processes and errors. This proves to be an already non-trivial task for the four-segment quantum repeaters as considered here. In principle, the state space of our MDP modeling the quantum repeater is infinite. By setting a maximum accumulated storage time $t_m$, which will be further defined in Sec. \ref{sec:modeling-a-multi-segment-quantum-repeater}, one could limit the size of the state space to $t_m^9$ for a four-segment quantum repeater. Thus, even when assuming $t_m$ as low as $t_m=10$ the number of states is $10^9$. Taking into account that there are at least two possible actions for any relevant state of the MDP and at least two possible transitions for each action, this lower estimation illustrates that analytical approaches are infeasible to solve this optimization problem. The present work can be understood as a proof of concept of a DRL-based optimization method for four-segment repeaters that is equally applicable to larger repeater chains where we expect it to be even more powerful compared with the standard approaches.

Ultimately, we find policies for the quantum memory treatment, which outperform the naive approaches used in the simulations. Therefore we demonstrate a successful proof-of-concept application of a DRL approach in a first step towards solving complex optimization tasks in quantum networks. The paper is organized as follows. In Sec. \ref{sec:modeling-a-multi-segment-quantum-repeater}, we will present our abstract model of a multi-segment quantum repeater chain. Then, in Sec. \ref{sec:simulating-key-distribution-based-on-quantum-repeaters}, we will describe and discuss the results of simulating four-segment quantum repeaters using this model. In Sec. \ref{sec:deep-reinforcement-learning}, we present our DRL approach and its results, especially in comparison with the simulations without DRL. Finally in Sec. \ref{sec:conclusion}, we conclude with a summary and conclusion, followed by several appendices giving additional technical details.

\section{Quantum repeater model}
\label{sec:modeling-a-multi-segment-quantum-repeater}
This section presents our model of a multi-segment quantum repeater including the relevant errors. We will use an MDP to describe the evolution of quantum states in a quantum repeater. We briefly introduce the secret key rate as a figure of merit to evaluate the performance of a quantum repeater. Lastly, we will discuss a memory cut-off policy, in which quantum states in the memories of a quantum repeater are discarded in order to improve the fidelity of the states distributed between the communicating parties.

\subsection{Physical model and parameters}
\label{sec:physical-model-and-parameters} 
Let us now introduce a simplified model of a multi-segment quantum repeater. This offers the possibility to obtain fairly general and conceptual results independent of any specific implementation.

\begin{figure}[tb]
	\fbox{\includegraphics[width=0.45\textwidth]{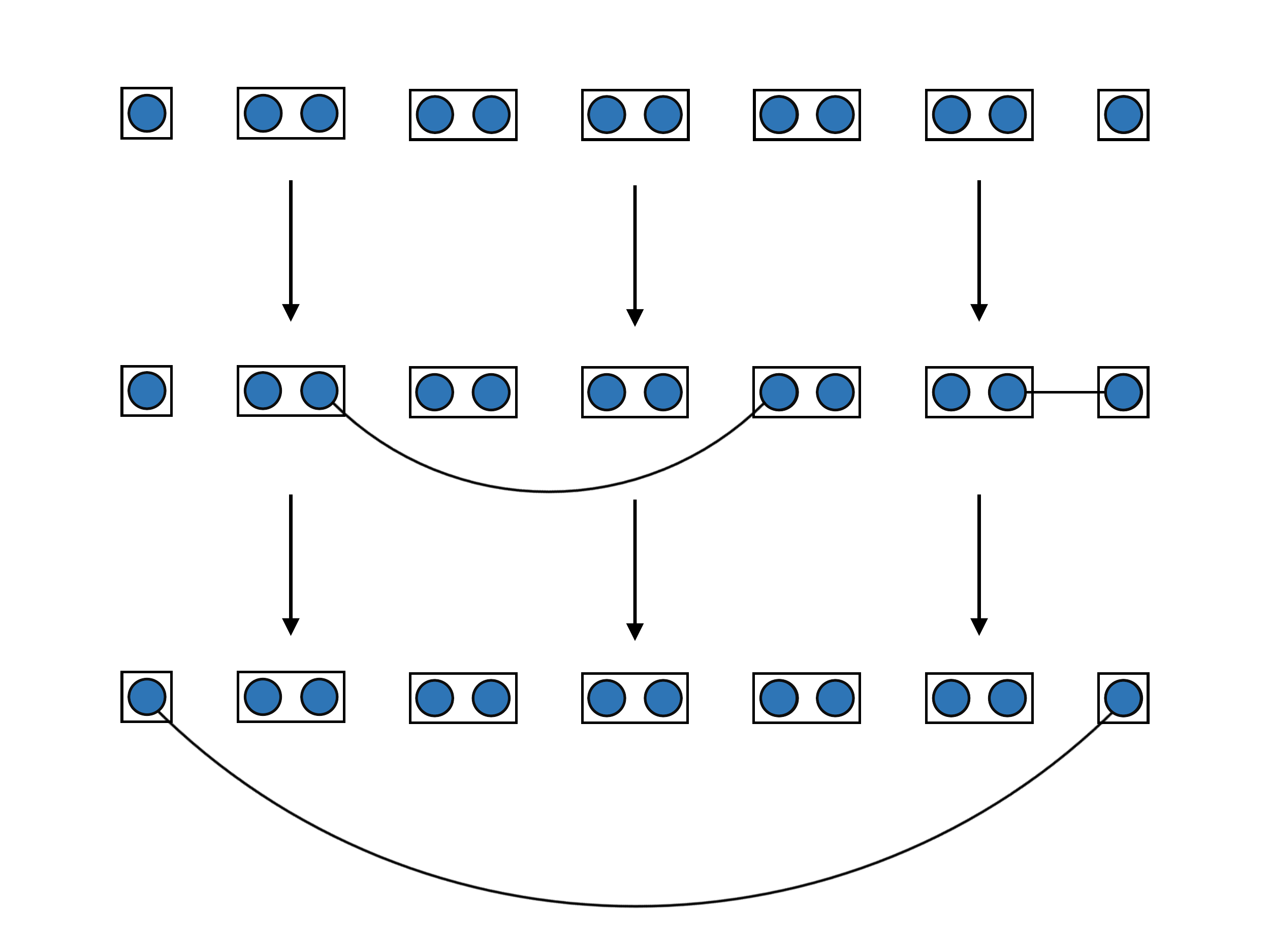}}
	\caption{\label{fig:quantum-repeater:n-segment-swapping} A multiple-segment quantum repeater chain. Each box depicts a quantum repeater node. Entangled two-qubit states are initially distributed within the segments. As soon as two-neighboring segments are ready, i.e. their memory pairs share a successfully distributed state, the entanglement swapping is performed. After subsequent swappings, synchronized via storage at the memory nodes, eventually an entangled state is shared between the two most outer nodes.}
\end{figure}

A simple, generic multi-segment quantum repeater chain is depicted in Fig. \ref{fig:quantum-repeater:n-segment-swapping}. Quantum repeaters are used to distribute entanglement between two distant parties by segmenting their connecting channel. Initial entanglement is generated in each segment, for instance, employing a source of entangled photon pairs at a node (or placed in the middle between two memory nodes) and sending one photon to each adjacent node. At each repeater node an entanglement swapping operation, which is essentially a Bell measurement for quantum teleportation, is performed on the memory qubits transferring the entanglement step by step over the entire distance to the communicating parties.

Throughout this work, quantum repeaters with the following properties are considered:
\begin{itemize}
\item The quantum repeater may in principle consist of an arbitrary number of segments. For simplicity, only one-dimensional concatenated segments are treated here, i.e. so-called quantum repeater chains (an extension to multidimensional repeater networks would also be possible with our methods, but this will not be demonstrated in this work). Eventually, we will put a particular emphasis on four-segment quantum repeaters.
\item Each repeater node contains one quantum memory for each adjacent node.
\item The repeaters are "clocked repeaters" where the clock times are determined by the classical communication times between nodes.
\item All errors of the implementation can be described as Pauli channels,
\begin{equation}
	\mathcal{N} \left( \rho \right) = \sum_{j=0}^{3} a_j P_j \rho P_j^\dagger,
	\label{eq:pauli-channel}
\end{equation}
where $\rho$ is the density operator of a quantum state, $P_j$ are the Pauli operators,
$
	\{ P_j \}_{ j \in \{ 0,1,2,3 \} } = \{ \mathds{1}, X, Z, XZ \},
$
and $a_j$ are real-valued, non-negative probabilities satisfying the normalization property
$
	\sum_{j=0}^3 a_j = 1.
$
\item Elements for active entanglement purification \cite{PhysRevA.53.2046,PhysRevLett.76.722} and more general quantum error correction are not included throughout. The only mechanism to suppress errors on the memories is a finite memory cut-off.
\end{itemize}

The parameters characterizing the implementation of a quantum repeater in our model are:
\begin{itemize}
\item $n$: The number of segments.
\item $L_0$: The length of one segment, which is the distance between adjacent repeater nodes.
\item $\nu_i \in [0,1]$: The probability of any error to occur in a given situation, in particular, the probability of a phase flip to occur on a quantum state stored in a memory.
\item $c$: The signal speed for classical communication between repeater nodes (typically, the signal speed in an optical fiber, which can be employed for transmitting both quantum and classical signals).
\item $L_{\text{att}}$: The attenuation length of the optical fibers for realizing the quantum channels between the repeater nodes.
\end{itemize}
The following, additional parameters can be obtained from those listed above:
\begin{itemize}
\item $\tau_0=\frac{L_0}{c}$: One round of quantum and classical communication, which is the time it takes for a node to send qubits and classical information to an adjacent node.
\item $\tau_c$: The memory coherence time for a bipartite quantum state, see Eq. (\ref{eq:dephasing}). It corresponds to half of the commonly used parameter of the single-qubit dephasing channel for the quantum memories.
\item $\eta = e^{-\frac{L_0}{L_{\text{att}}}}$: The transmissivity of the optical fibers connecting the repeater nodes, inducing an exponential photon loss with distance.
\item $p = p_x \eta$: The probability of generating an initial two-qubit entangled state in a segment in one time step. The parameter $p_x$ incorporates the photon creation efficiency of the spin-photon or photon pair source, fiber channel in- and out-coupling and detector efficiencies, as well as memory write-in efficiencies.

The attenuation length $L_{\text{att}}$ is a physical parameter of the optical fiber dependent on the employed wavelength of the transmitted photons. The preferred wavelength is that of telecommunication ($1.55 \text{ } \mu \text{m}$, potentially requiring wavelength conversions) and a typical value for $L_{\text{att}}$ is 22 km.

Note that the particular form of the transmissivity $\eta$ relies on the assumption of constant photon losses, which is the case for optical fibers and this would have to be changed for different channel implementations (such as fluctuating loss in a free-space channel).

\end{itemize}

\subsection{Markov decision process}
\label{sec:mdp}
For the purpose of adequately describing a quantum repeater, we designed an MDP to model the propagation and evolution of the quantum states stored in the quantum memories of a multi-segment quantum repeater in our physical model.

An MDP is defined as a tuple $(S,A,P,R)$ where
\begin{itemize}
\item $S$ is a set of states of the environment,
\item $A$ is a set of actions performed by the agent on the environment,
\item $P : S \times A \times S \rightarrow [0,1]$  is the transition probability in a time step between states dependent on the applied action in this time step,
\item $R : S \times A \times S \rightarrow [0,1]$ is the immediate reward received from a transition.
\end{itemize}

In practice, besides photon losses, Pauli channels include the most common errors in quantum repeaters and networks \cite{PhysRevLett.81.5932}. This is fortunate from the mathematical perspective, since Pauli channels commute with the entanglement swapping operations \cite{PhysRevA.102.042614,PhysRevLett.87.267901}. Thus, in the case of memory dephasing, which is a Pauli channel, the number of accumulated error operations is proportional to the accumulated storage time, i.e. additively propagating the storage times of quantum states through the swapping operations gives the correct accumulated dephasing. Therefore, instead of rigorously calculating the density matrix of every intermediate quantum state, it is sufficient to treat the undisturbed quantum states that are only subject to channel loss and count the accumulation of errors separately. Then all these accumulated errors can be applied to the final quantum state distributed between the communicating parties. Hence, the states of the MDP can be encoded in a triangular matrix where each entry corresponds to a pair of repeater nodes storing the accumulated errors of the bipartite quantum state. One time step of the MDP is the time it takes to send quantum and classical information between adjacent repeater nodes (corresponding to $\tau_0$ in our physical model). Note that in our treatment here we only consider the simplest, still to some extent idealized scenario with only two effects determining the final rates: channel loss and memory dephasing. This channel-loss-and-memory-dephasing-only model is similar to that of Ref. \cite{https://doi.org/10.1002/qute.201900141}, however, here extended from two to four repeater segments. Additional experimental (error) parameters have been included in the analytical treatment of Ref. \cite{https://doi.org/10.48550/arxiv.2203.10318}. As opposed to the present work, Ref. \cite{https://doi.org/10.48550/arxiv.2203.10318} has no focus on optimizing the memory cut-off (for a discussion on other existing works and approaches for optimizing the memory cut-off, see Sec. \ref{sec:overview-of-exising-approaches}).

The set of possible actions of the agent consists of arbitrary combinations of swapping and discarding operations. The action applied in one time step may perform entanglement swapping on any subset of repeater stations and discard any subset of the stored quantum states. In the event that the two outer nodes of the repeater share an entangled state, this state is discarded and its fidelity is returned as a reward. A detailed description of the MDP can be found in App. \ref{sec:markov-decision-process-modeling-a-quantum-repeater}.

\subsection{Quantum key distribution - secret key rate}
The secret key rate $R$ is a suitable measure to evaluate the performance of a quantum repeater, as it combines the relevant properties, namely quantum state fidelity and raw transmission rate, into one convenient figure of merit. Furthermore, long-range QKD is one of the main applications motivating the development of quantum repeaters. The secret key rate is defined as the amount of secret bits distributed between two communicating parties, commonly denoted Alice and Bob, in bits per time and can be written as
\begin{equation}
	R = Y \cdot r ,
\end{equation}

where $Y$ is the raw rate and $r$ the secret key fraction. The raw rate is the amount of raw (quantum) bits distributed between Alice and Bob per time and the secret key fraction is the potential amount of secret key that can be extracted per raw (quantum) bit via post-processing.

Throughout this work the Lo-Chau BB84 \cite{BENNETT20147,Lo2005} protocol is used to determine the secret key rates. The asymptotic secret key rate of the Lo-Chau BB84 protocol, see for example Ref. \cite{pir2019advances}, reads as

\begin{equation}
	R_{\text{BB84}} = Y \cdot r_{\text{BB84}} \left( e_1 , e_2 \right) ,
	\label{eq:BB84}
\end{equation}
with the secret key fraction
\begin{equation}
	r_{\text{BB84}} \left( e_1 , e_2 \right) = 1 - h\left( e_1 \right) - h\left( e_2 \right),
	\label{eq:skfraction}
\end{equation}
where $h$ is the binary entropy function defined on the interval $\left[ 0,1 \right]$ as
\begin{equation}
	h\left( p \right) = - p \text{ log } p - ( 1 - p ) \text{ log} \left( 1 - p \right),
\end{equation}
where the logarithm is taken to the base two.

The quantum bit error rates $e_1$ and $e_2$ are the probabilities of non-coinciding measurements performed by the two parties on the shared data qubits in the respective coinciding basis. Commonly, the Pauli $X$ and Pauli $Z$ bases are chosen. In a realistic scenario, Alice and Bob have to estimate the bit error rates on a finite set of test data. In the proposal of Ref. \cite{Lo2005} it was shown that the number of test bits can be chosen in the order of $\Omega ( \text{log } k)$, where $k$ is the length of the final key, while still achieving unconditional security. Thus, the fraction of test data can be chosen asymptotically close to zero.

The concrete derivation of the asymptotic secret key rate for our physical model, including only fiber losses and memory dephasing (see later Eq. (\ref{eq:dephasing})), can be found in App. \ref{sec:calculation-of-the-secret-key-rates} and it takes the simple form

\begin{equation}
	R_{\text{BB84}} = Y \cdot \left( 1 - h\left( \frac{1}{2} \left( 1 - \mathds{E} \left[ e^{- \frac{t}{\tau_c}} \right] \right) \right) \right),
	\label{eq:detailed-skr}
\end{equation}
where $\tau_c$ is the coherence time of the quantum memory for a bipartite quantum state and $t$ is the storage time of the quantum state.

\subsection{Memory dephasing - cut-off and swapping strategies}
\label{sec:dephasing-strategies}

\subsubsection{Overview of existing approaches}
\label{sec:overview-of-exising-approaches}
One of the most significant error sources in practical implementations of memory-based quantum repeaters is the degradation of quantum states in the memories. The distribution of entanglement via a quantum repeater typically is a highly probabilistic process. Thus, the level of degradation caused by memory dephasing can become very large, with some nodes having to wait for others for so long that entanglement can hardly be preserved.

As already mentioned in the introduction, limiting the time during which the quantum states are stored is a common strategy to improve the fidelity of the distributed quantum states. This concept adapted to quantum repeaters based on imperfect memories was first introduced in Ref. \cite{PhysRevLett.98.060502}. In terms of resources, this is the least expensive approach to suppress memory errors compared to entanglement purification strategies relying on the distribution of additional entangled-state copies and extra classical communication rounds or, alternatively, strategies based on the implementation of more complicated quantum error correction codes. Again, the processes in a quantum repeater are highly probabilistic and computing the corresponding probability distributions for complex repeater schemes quickly becomes infeasible. In recent years, significant research has been devoted to this and results under various simplifying assumptions have been reported. In order to put our work and results in context we shall give a brief overview of some of the existing literature.

The distributed entangled quantum states in a two-segment repeater including cut-offs were thoroughly analyzed in the presence of various error models in Refs. \cite{Rozp_dek_2018,PhysRevA.99.052330,https://doi.org/10.1002/qute.201900141,PhysRevA.102.042614}. Reference \cite{PhysRevA.102.042614} also considers more than two segments, however, adapted to specific distribution protocols. In Ref. \cite{praxmeyer2013reposition} a closed, compact formula is derived to efficiently compute the raw rate of a quantum repeater with an arbitrary number of segments, including a memory cut-off, with the constraint of deterministic entanglement swapping operations and the (practically suboptimal) assumption that all swappings are performed at the end. In Ref. \cite{PhysRevA.100.032322} an algorithm based on Markov chains and solving linear-equation systems is presented which exactly computes the average waiting time of quantum repeater chains with an, in principle, arbitrary number of segments including various swapping strategies. For up to four segments exact rate formulas are given. However, this algorithm is practically limited as its runtime in $\mathcal{O}(c^n)$ growth rather quickly with the number of segments $n$ and the cut-off $c$. In \cite{PhysRevLett.128.150502} swapping strategies in the presence of non-deterministic swapping operations are optimized for the best possible raw rate, but no cut-off was included.

In a more numerical approach, in Ref. \cite{Jiang17291} an efficient optimization over entanglement distribution schemes using dynamical programming is proposed under the assumption that there is no time-dependent decoherence in the quantum memories. The algorithm recursively solves larger repeater chains by dissecting them into smaller sub-chains which are optimized requiring the idealizing simplification that the sub-processes finish at the average time. In Ref. \cite{Santra_2019}, the optimal memory buffer maximizing the rate of distillable entanglement of the average state at all nesting levels in a doubling repeater scheme is computed. In another work, in Ref. \cite{8972391} algorithms to compute the probability distribution of the fidelity and waiting time for the first distributed entangled state in a quantum repeater and a numerical optimization over a cut-off is presented. Moreover, in Ref. \cite{khatri2020policies}, it was shown that the optimal policy maximizing the accumulated fidelity in a multiplexed repeater segment can, in the finite-horizon setting, be computed via a dynamical programming algorithm. In Ref. \cite{PhysRevA.103.032610}, a heuristic algorithm was developed which optimizes quantum repeater schemes in order to minimize the generation time of an entangled state between communicating parties for a fixed minimum success probability and fidelity. Their schemes also include memory buffers and entanglement distillation. To overcome the computational complexity the schemes were restricted to those that succeed at all levels near-deterministically. This is enforced by repeating all probabilistic processes a sufficient number of times to ensure a high probability of at least a single success. In Ref. \cite{Ferreira_da_Silva_2021}, a genetic algorithm was presented and applied to NetSquid simulations \cite{Coopmans2021} providing insights into the necessary hardware parameters for viable quantum repeaters. Recently, in Ref. \cite{https://doi.org/10.48550/arxiv.2203.10318}, an exact rate analysis for quantum repeaters including experimental errors was presented. However, the cut-off was included primarily in a sequential repeater scheme and no optimization of the cut-off was made. Reference \cite{9495278}, as most relevant to the present work, was already discussed in the introduction.

As was stated in the introduction, our policies can dynamically adapt to the state of the repeater and decide for any individual quantum state if it should be discarded or not. Thus, we include an extended toolbox which offers the possibility of more sophisticated and, as we will present in Sec. \ref{sec:drl-results}, better policies than previous approaches. In order to handle the complexity introduced by this generalization we use a DRL algorithm which is also novel in the optimization of quantum repeaters. We will distinguish the two policies by referring to the simpler policy as the cut-off policy and to the more sophistcated, better policy as learned policies.

\subsubsection{Cut-off model and policies in this work}
In this paper, the cut-off policy is defined such that any quantum state whose accumulated storage time exceeds a chosen cut-off value $c$ will be discarded. Accumulated storage time here refers to the storage time that propagates additively through the swapping operations leading to the final quantum state shared between the most distant stations. If, for example, a state is stored for $n_1$ time steps and swapped with a state stored for $n_2$ time steps, the state after the swapping has an accumulated storage time of $n = n_1 + n_2$ time steps. Note that this differs from other common definitions of the cut-off, which usually only use the time the state is stored for in the current nesting level. However, as was explained in Sec. \ref{sec:mdp}, our accumulated storage time does, in fact, correctly describe the propagated quantum state. This substantiates the choice of an accumulated cut-off, since it determines the quality of a quantum state more accurately than when only the storage time in one nesting level is taken into account. Also, note that a cut-off per nesting level was employed in Ref. \cite{9495278} where it was found that for the presented example parameters a non-uniform cut-off, i.e. a different cut-off for each nesting level, does not yield a significant improvement of the secret key rate.

The cut-off policy as defined in our work, based on an accumulated storage time, actually ignores the fact that the information on which the decision to discard quantum states is based on might not be directly available at the node performing the operation at the corresponding time. Our basic assumption that the necessary information is indeed available is not obvious and may even seem unreasonable. Also, an argument that this serves as a bound is invalid, as classical communication is not an obstacle one could possibly circumvent.
First, one should note that this does not physically contradict anything of how the quantum repeater is operated and functions. The only but clearly idealizing element of this is that decisions for the controlled part of the process, namely the discarding of the quantum states, are generally based on the entire state of the MDP. This means decisions of the policy performing an action in one node might be based on information that cannot be possibly accessible in this particular node. In the context of examining fully realistic quantum repeaters, all necessary classical communication must be taken into account. However, in this work, in order to allow for optimal comparability between unlearned and learned policies we choose these to be all-knowing to give the algorithm at any time complete information to discover new policies. Therefore, imposing the same conditions on the unlearned policies, simulated in the following section, as those imposed on the learned policies, described later, ensures that any advantage of the learned policies must be based on their better strategy and will not be based on any better assumptions.

Next, we want to briefly discuss swapping strategies. As a simplification, throughout this work, we assume the entanglement swapping to be error-free and deterministic. In the case that memory dephasing is the only error acting on the quantum states, the ideal swapping strategy is to perform the Bell measurement at any repeater node immediately when both adjacent segments have successfully distributed entanglement. This follows from the fact that a swapping operation reduces the number of bipartite quantum states that are simultaneously stored and thus the accumulated dephasing (for a rigorous and more mathematical proof of this statement, together with a systematic formalism to exactly calculate the secret key rate in such protocols that "swap as soon as possible", see Ref. \cite{https://doi.org/10.48550/arxiv.2203.10318}).

All schemes simulated in this work are what we call overlapping schemes. This means that, opposed to other common analysis which often only considers the first distributed state, we consider the generation of many distributed states in an overlapping fashion. This further means that in the lower nesting levels states will be continued to be generated for future processing while other memories are still occupied for the next distributed state. In the four-segment case treated in this work this might not have a significant effect on the performance of the repeater, as the only case where this is actually beneficial is when an entangled state over three segments is shared. In this case, the segment in the middle of this connection can again start entanglement generation attempts.

\section{Simulating key distribution based on quantum repeaters}
\label{sec:simulating-key-distribution-based-on-quantum-repeaters}
In this section, the cut-off policy is examined via simulations with respect to its effect on the secret key rates. The primary purpose of this is to provide benchmarks of achievable secret key rates to be surpassed, for the policies later considered with the RL algorithm in Sec. \ref{sec:deep-reinforcement-learning}. Before presenting in detail the results of the simulation, we shall briefly describe the specific errors included in the simulation, extending the discussion on our physical model and errors from Sec. \ref{sec:physical-model-and-parameters}.

\subsection{Errors and imperfections}
A realistic quantum repeater is a complex system with numerous parameters. The more realistic effects are included in a simulation, the more meaningful the results become regarding the assessment of a practical quantum repeater. On the other hand, excluding some sources of errors allows for a more focused view of the most important ones. Another way to look at this is that the results considering fewer errors serve as upper bounds for what is achievable.

The focus of this work is on strategies to counteract the dephasing of quantum memories. Therefore the only two imperfections that we choose to include are the dephasing of the quantum memories and the finite transmissivity of the quantum channel (while without the latter quantum repeaters would be pointless).

We model the degradation of the quantum states that are stored in the memories as a dephasing channel,
\begin{equation}
	\mathcal{N}_Z \left( \rho \right) = \left( 1 - \nu \right) \rho + \nu Z \rho Z ,
	\label{eq:dephasing-channel}
\end{equation}
which we further specify as exponential decay at time $t$,
\begin{equation}
	\mathcal{N}_Z \left( \rho , t \right) = \frac{1}{2} \left( 1 + e^{- \frac{t}{2 \tau_c}} \right) \rho + \frac{1}{2} \left( 1 - e^{-\frac{t}{2 \tau_c}} \right) Z \rho Z,
	\label{eq:dephasing}
\end{equation}
where $Z$ is a Pauli operator and $\tau_c$ is the coherence time of the quantum memory \cite{Rozp_dek_2018}. Note that the dephasing channel as defined in Eq. (\ref{eq:dephasing}) is a single-qubit channel. In a quantum repeater, both qubits of a stored entangled bipartite state are subject to this error channel (individually, i.e. locally and independently) prior to entanglement swapping. This is also the reason for the factor $2$ multiplied to the coherence time $\tau_c$, which was defined as the coherence time for a bipartite quantum state. The second imperfection included is the photon loss in the optical fibers, which was already described in Sec. \ref{sec:physical-model-and-parameters}.

\subsection{Results}
\label{sec:simulations:results}
This section presents the results of our numerical simulations. In the simulations, the secret key rates for a BB84 protocol are computed. A discussion of the required formulas can be found in App. \ref{sec:calculation-of-the-secret-key-rates}. Two-segment quantum repeaters have already been analyzed rather thoroughly \cite{Rozp_dek_2018,PhysRevA.99.052330} and, being structurally simple, do not offer extra insights for our treatment. In this work, we choose to consider the class of four-segment repeaters. This is partially motivated by the existing literature, which often considers doubling the number of repeater segments, making four segments the logical next step beyond two. Here it is also worth noting that the more segments are included, the less feasible it can become in our model without entanglement purification and error correction that meaningful practical or even non-zero secret key rates can be obtained \cite{https://doi.org/10.48550/arxiv.2203.10318}. For repeaters of a larger scale such extra tools will have to be included.

For the attenuation length $L_{\text{att}}$ we choose 22 km which is today's experimental standard for the telecom wavelength of 1550nm \cite{RevModPhys.81.1301,Agrawal_1997}. The speed of classical communication is assumed to be $2 \cdot 10^8 \text{m/s}$, which corresponds to the typical value in optical fibers. The segment lengths are chosen within a commonly considered range of 20 - 70 km, and the coherence time of the memories is assumed to be within the scope of today's experimental possibilities \cite{https://doi.org/10.1002/qute.201900141}.

\subsubsection{Uncertainties}
In the simulations, since overall minimal uncertainties were achieved, 3-$\sigma$ confidence intervals ($\approx$ 99,73\%) are displayed in all plots with uncertainty bars, instead of the more commonly chosen $\sigma$ intervals ($\approx$ 68,27\%). However, note that the uncertainty bars are still small enough for many data points not to be easily visible, as they can be smaller than the displayed points of the data. One should not be misled by their width in the horizontal direction, since this axis corresponds to a discrete quantity and does not display an uncertainty. The style of displaying the uncertainties was chosen to increase their visibility. Their height corresponds to the 3-$\sigma$ confidence interval of the secret key rate.

It is also worth noting that the uncertainties get larger for a sparser entanglement distribution in time, thus for longer segment length and smaller cut-off parameters. The reason is that the sparser the received quantum states are, the larger the required sample size is to obtain an equally good estimation of the average fidelity. Hence, the precision of some values with more significant uncertainties is limited by the same computational time needed to simulate larger sample sizes.

\subsubsection{Quantum repeater without cut-off}
\begin{figure}[tb]
\makebox[0.45\textwidth][c]{
	\includegraphics[width=0.264\textwidth]{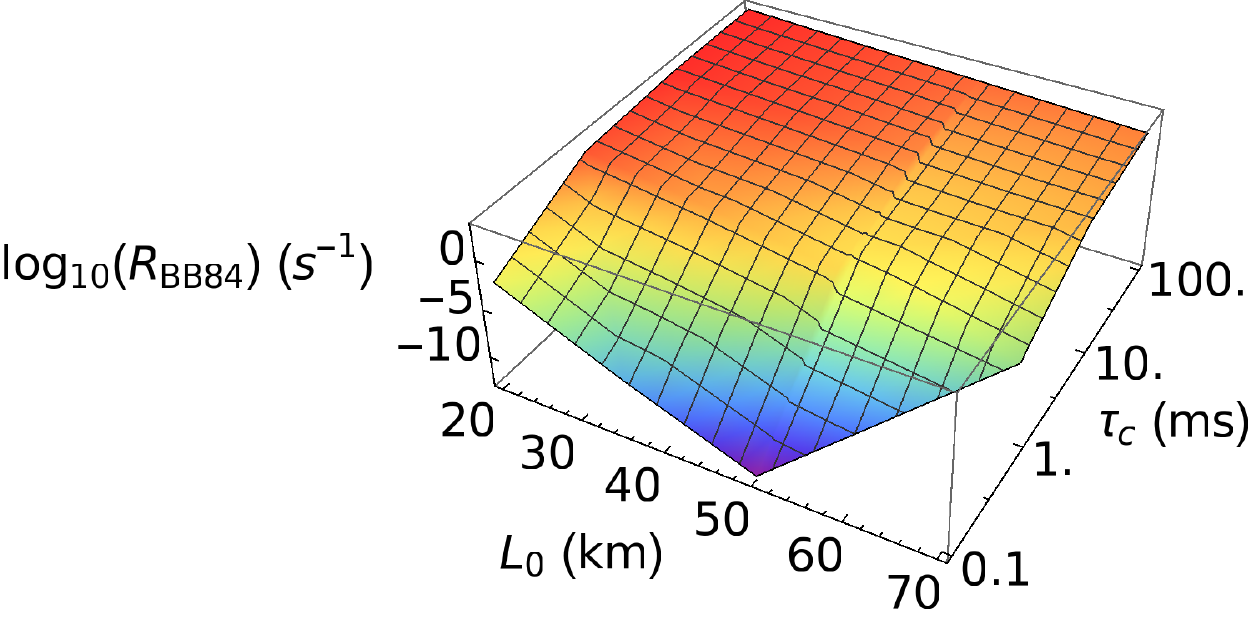}
	\includegraphics[width=0.176\textwidth]{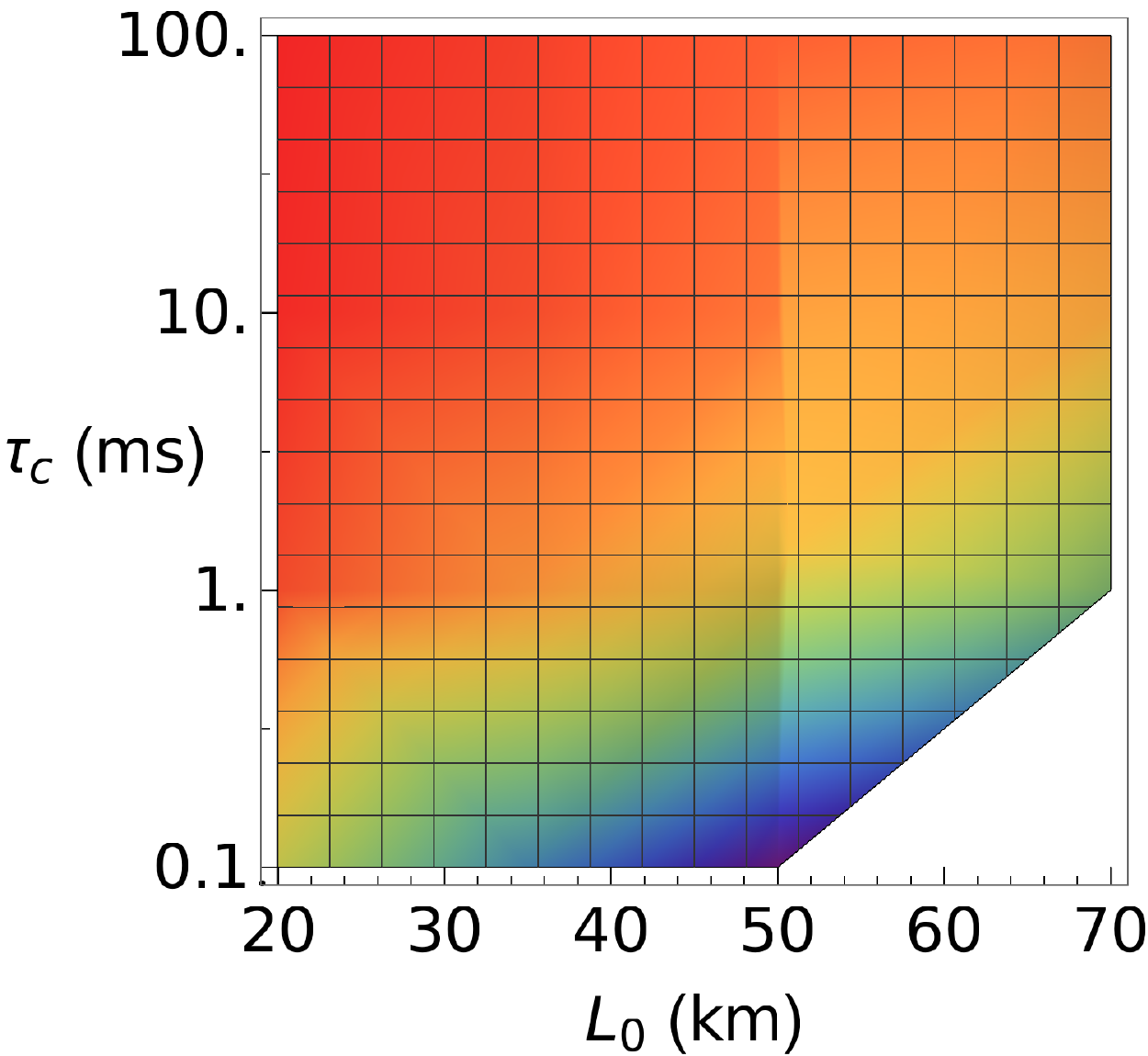}
	\hspace{30pt}
}
\makebox[0.45\textwidth][c]{
\begin{minipage}{\baselineskip}
	\hfill\vspace{\baselineskip}
	$R_{\text{BB84}}\text{  }$($s^{-1}$)
\end{minipage}
}
\makebox[0.45\textwidth][c]{
	\includegraphics[width=0.264\textwidth]{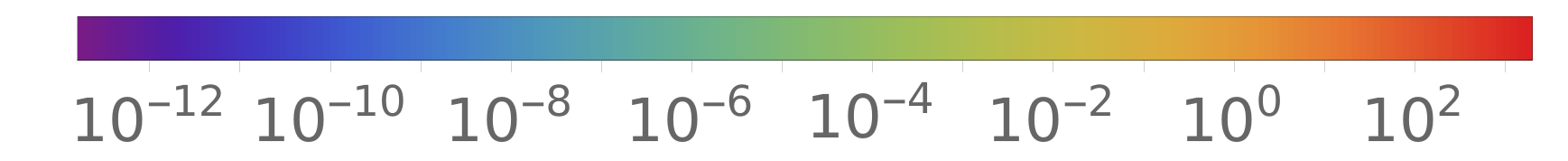}
}
	\caption{\label{fig:nocutoff}Secret key rates of four-segment quantum repeaters without memory cut-off. The secret key rate is plotted in secret bits per second dependent on the segment length $L_0$ and the memory coherence time for a bipartite quantum state $\tau_c$. The secret key rate and the coherence time of the memories are displayed on a logarithmic scale.}
\end{figure}

In this subsection, the secret key rate of quantum repeaters that are not controlled by a policy including memory cut-off is examined. The results are presented in Fig. \ref{fig:nocutoff}.

In order to deepen our understanding, we now further elaborate some implications of the parameters. One should note that since the classical communication time $\tau_0 = \frac{L_0}{c}$ scales linearly with the segment length $L_0$, a larger segment length $L_0$ does not only decrease the transmissivity of the channel, but it also decreases the effective coherence time per channel use. Furthermore, an increase in classical communication time results in lower raw rates per second for identical raw rates per channel use, thus reducing the efficiency of a channel use. However, these drawbacks are reasonable taking into account that the overall communication distance increases with the segment length, thus fundamentally lowering the transmission rate.

The results obtained in our simulations are what one would intuitively expect, as the secret key rate increases with smaller repeater segments and better coherence time of the memories. The secret key rate saturates for sufficiently large coherence times, because the fidelity of the distributed quantum states approaches unity, and it drops to zero for short coherence times when the fidelity approaches a value of half. With longer segment length the maximum secret key rate decreases. The larger the segment length, the higher are the requirements on the quantum memories to achieve reasonable secret key rates. The increase of the secret key rates with the quality of the memories is steeper for worse memories and gets less significant, as the secret key rate saturates towards perfect memories. The decrease of the secret key rate with increasing segment distance is significantly steeper for worse memories.

\subsubsection{Quantum repeater with cut-off}

\begin{figure}[tb]
\makebox[0.45\textwidth][c]{
	\includegraphics[width=\plotsize\textwidth]{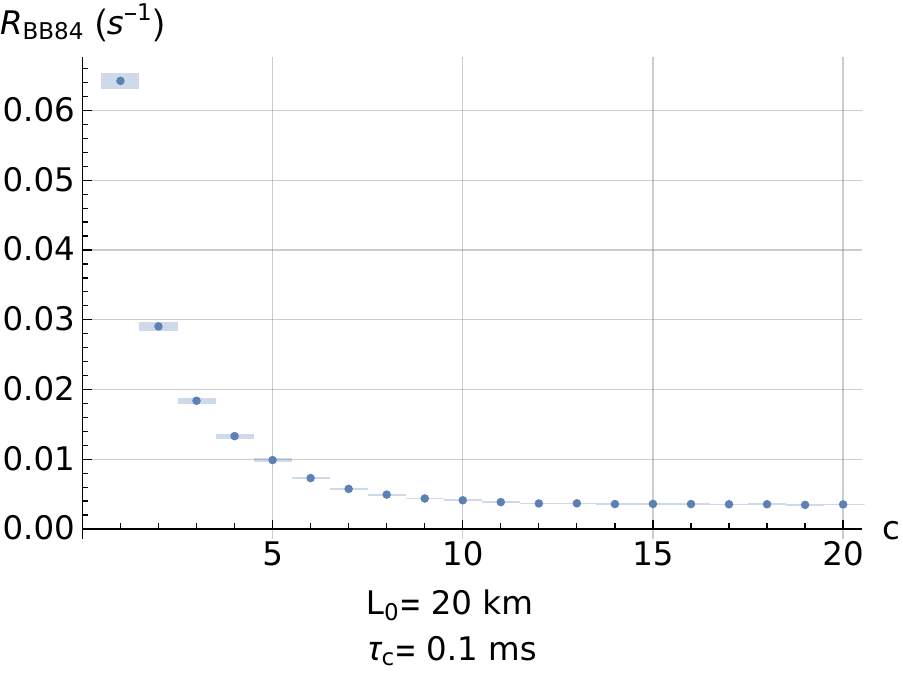}
	\includegraphics[width=\plotsize\textwidth]{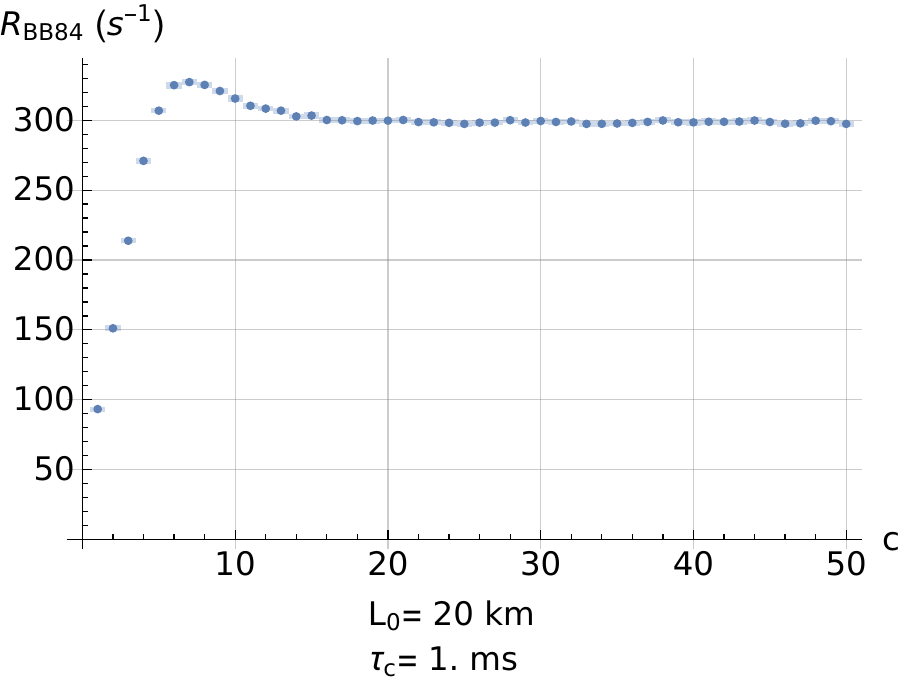}
}
\makebox[0.45\textwidth][c]{
\begin{minipage}{\baselineskip}
	\hfill\vspace{\baselineskip}
\end{minipage}
}
\makebox[0.45\textwidth][c]{
	\includegraphics[width=\plotsize\textwidth]{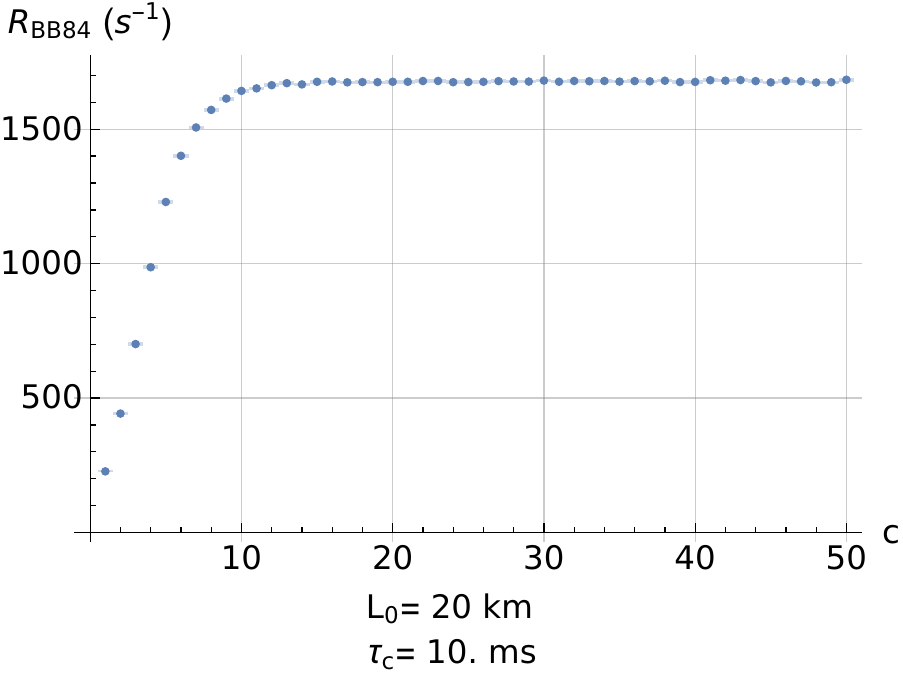}
	\includegraphics[width=\plotsize\textwidth]{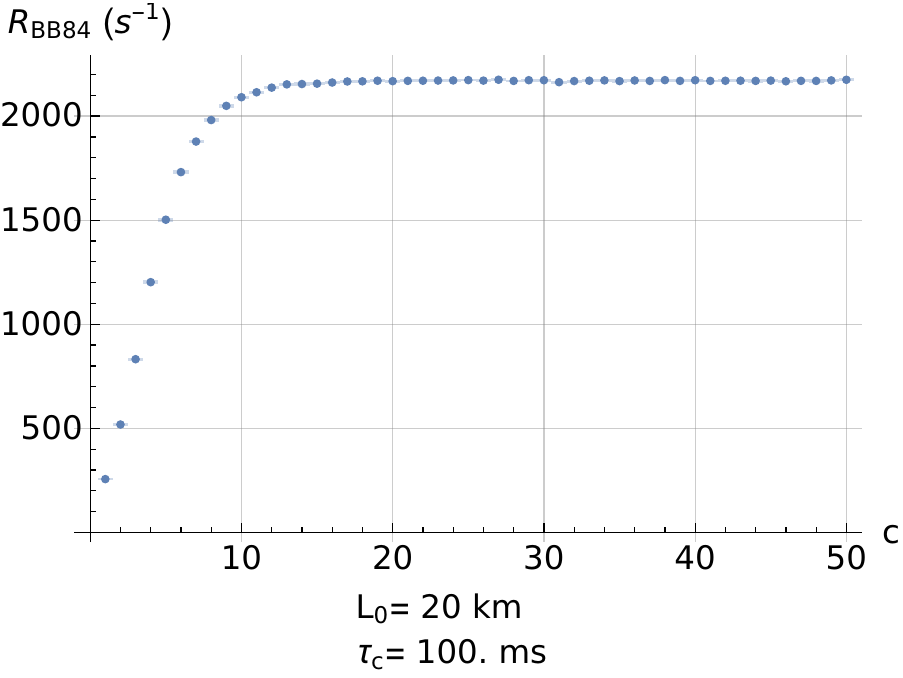}
}
	\caption{\label{fig:cutoff-20}Secret key rates of four-segment quantum repeaters in secret bits per second dependent on the memory cut-off parameter $c$ for segment length $L_0 = 20 \text{ km}$ and different coherence times of the memories for a bipartite quantum state $\tau_c$. Plotted with 3-$\sigma$ confidence intervals.}
\end{figure}

\begin{figure}[tb]
\makebox[0.45\textwidth][c]{
	\includegraphics[width=\plotsize\textwidth]{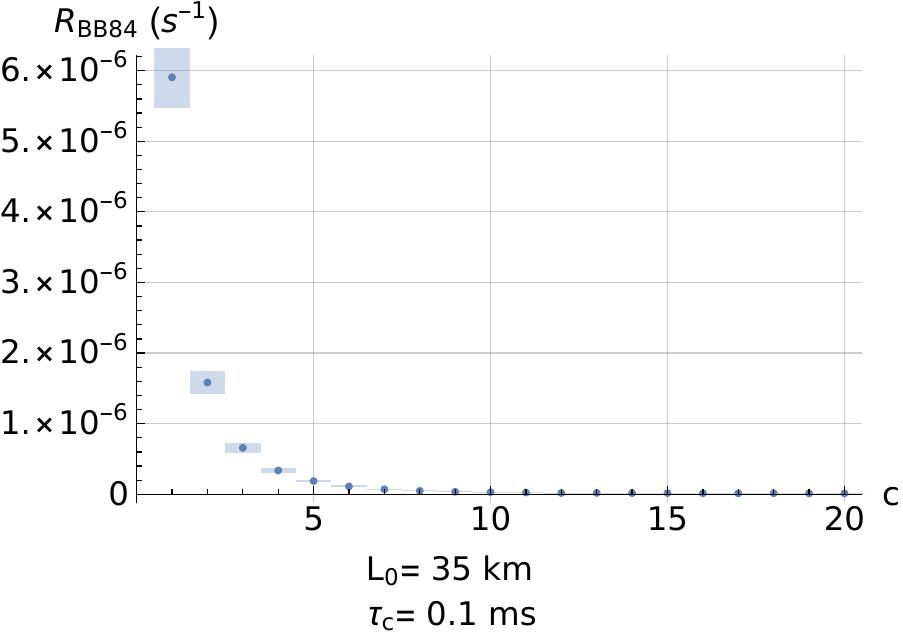}
	\includegraphics[width=\plotsize\textwidth]{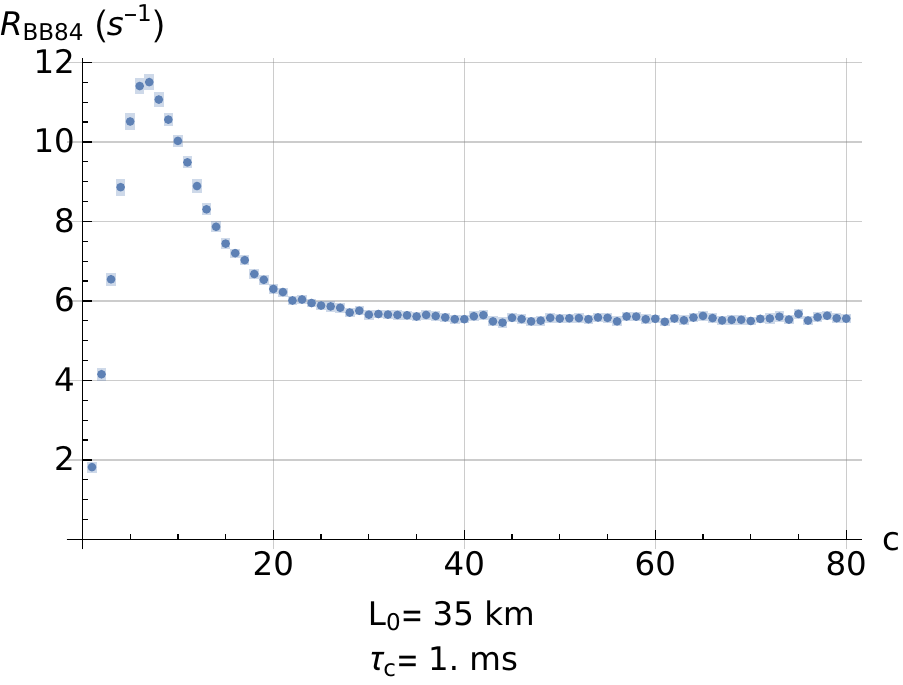}
}
\makebox[0.45\textwidth][c]{
\begin{minipage}{\baselineskip}
	\hfill\vspace{\baselineskip}
\end{minipage}
}
\makebox[0.45\textwidth][c]{
	\includegraphics[width=\plotsize\textwidth]{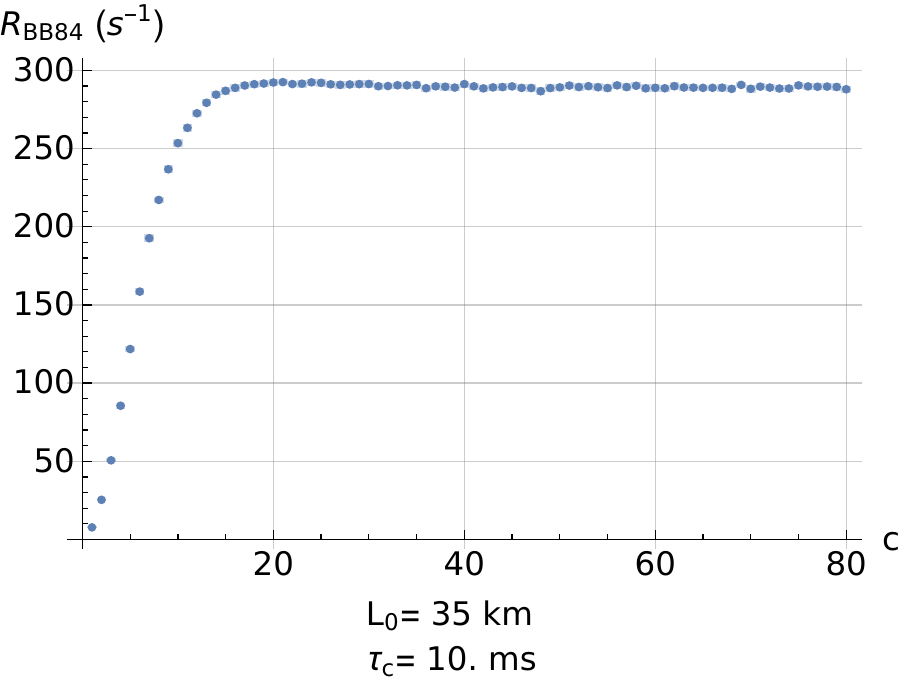}
	\includegraphics[width=\plotsize\textwidth]{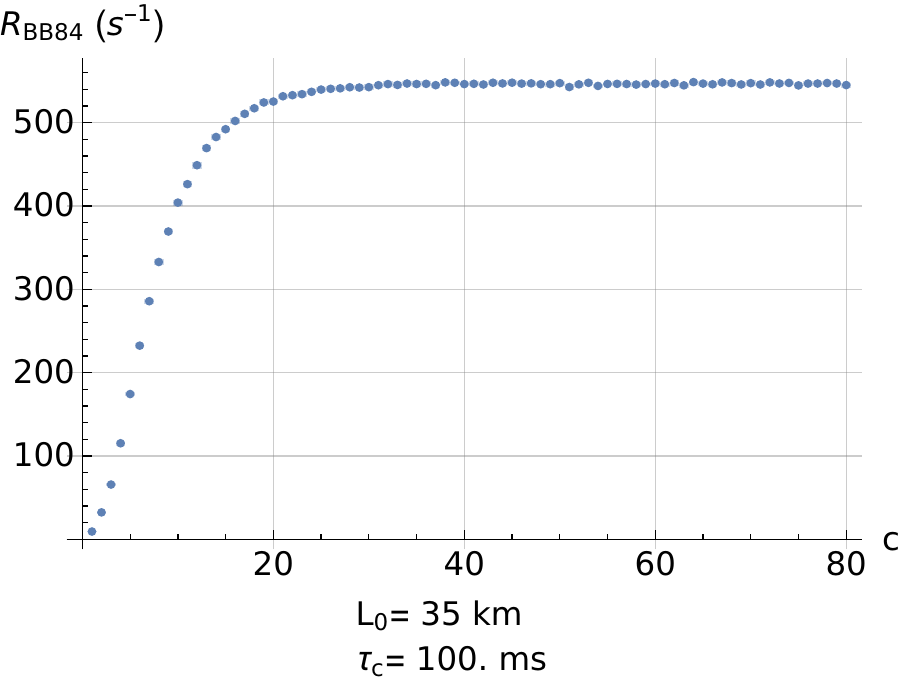}
}
	\caption{\label{fig:cutoff-35}Secret key rates of four-segment quantum repeaters in secret bits per second dependent on the memory cut-off parameter $c$ for segment length $L_0 = 35 \text{ km}$ and different coherence times of the memories for a bipartite quantum state $\tau_c$. Plotted with 3-$\sigma$ confidence intervals.}
\end{figure}

\begin{figure}[tb]
\makebox[0.45\textwidth][c]{
	\includegraphics[width=\plotsize\textwidth]{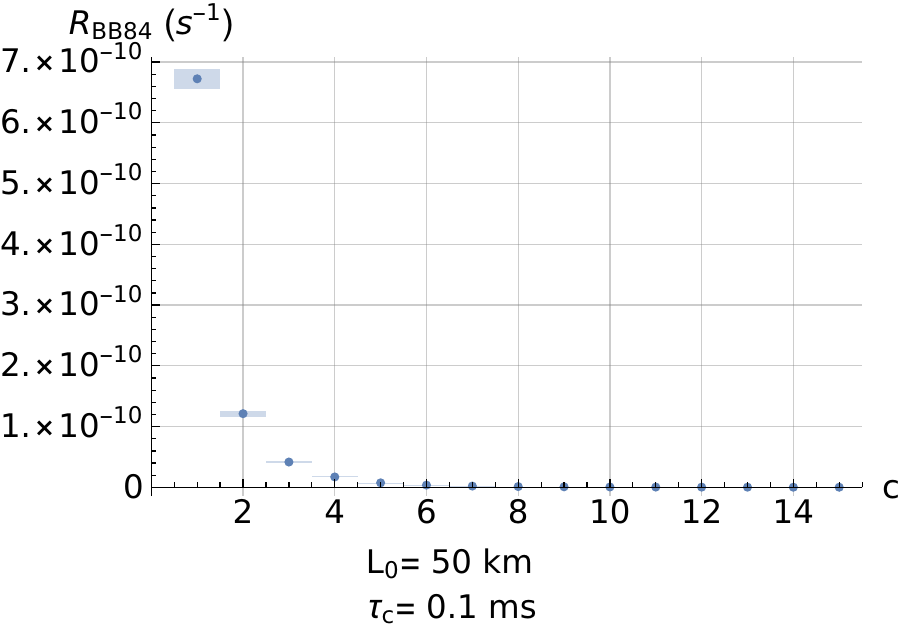}
	\includegraphics[width=\plotsize\textwidth]{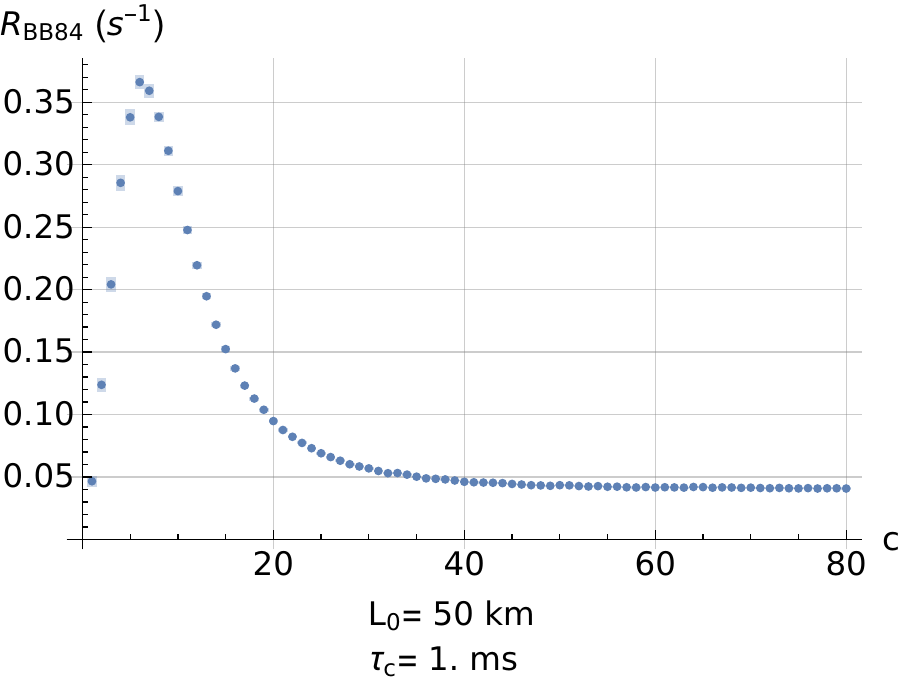}
}
\makebox[0.45\textwidth][c]{
\begin{minipage}{\baselineskip}
	\hfill\vspace{\baselineskip}
\end{minipage}
}
\makebox[0.45\textwidth][c]{
	\includegraphics[width=\plotsize\textwidth]{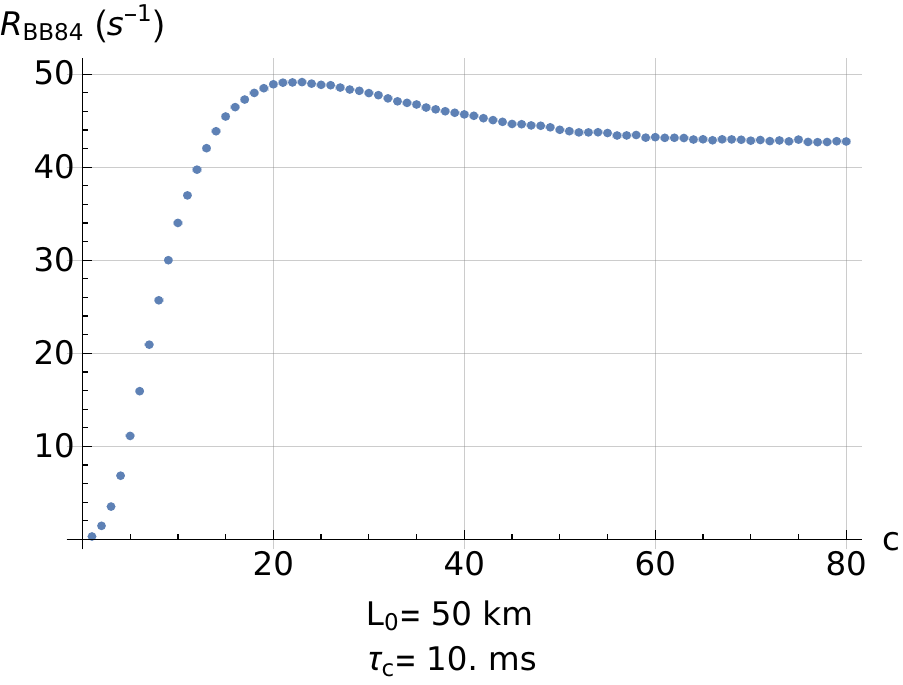}
	\includegraphics[width=\plotsize\textwidth]{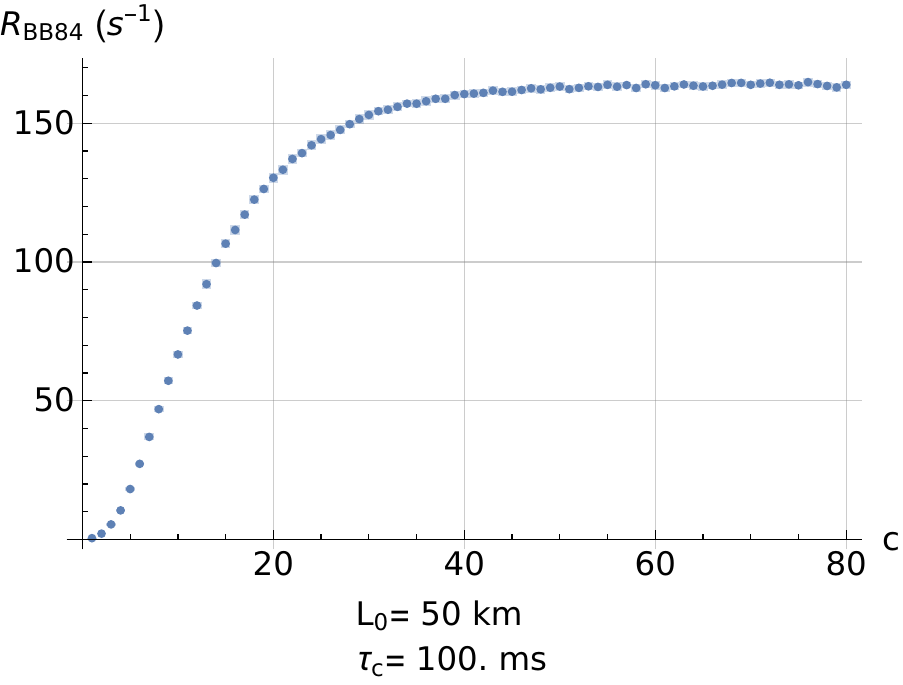}
}
	\caption{\label{fig:cutoff-50}Secret key rates of four-segment quantum repeaters in secret bits per second dependent on the memory cut-off parameter $c$ for segment length $L_0 = 50 \text{ km}$ and different coherence times of the memories for a bipartite quantum state $\tau_c$. Plotted with 3-$\sigma$ confidence intervals.}
\end{figure}

\begin{figure}[tb]
\makebox[0.45\textwidth][c]{
	\includegraphics[width=\plotsize\textwidth]{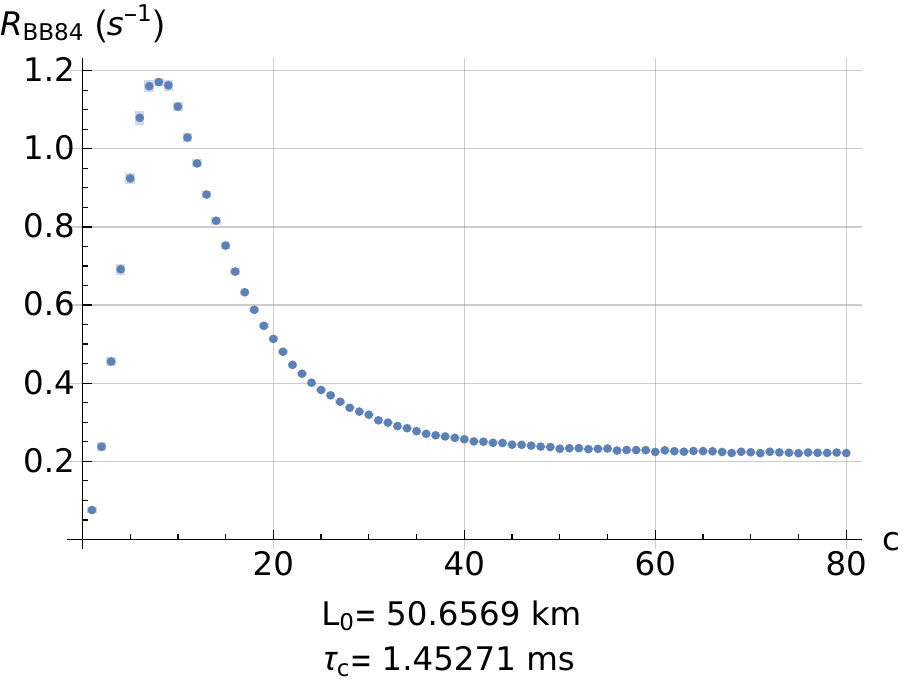}
}
	\caption{\label{fig:cutoff-50-b}Secret key rates of four-segment quantum repeaters in secret bits per second dependent on the memory cut-off parameter $c$ for segment length $L_0 = 50.6569 \text{ km}$ and coherence times for a bipartite quantum state $\tau_c = 1.45271 \text{ ms}$. Plotted with 3-$\sigma$ confidence intervals. \footnote{These values correspond to the probability to generate initial entanglement in one segment $p = 0.1$ and the dephasing probability in one time step $\nu = 0.08$.}}
\end{figure}

\begin{figure}[tb]
\makebox[0.45\textwidth][c]{
	\includegraphics[width=\plotsize\textwidth]{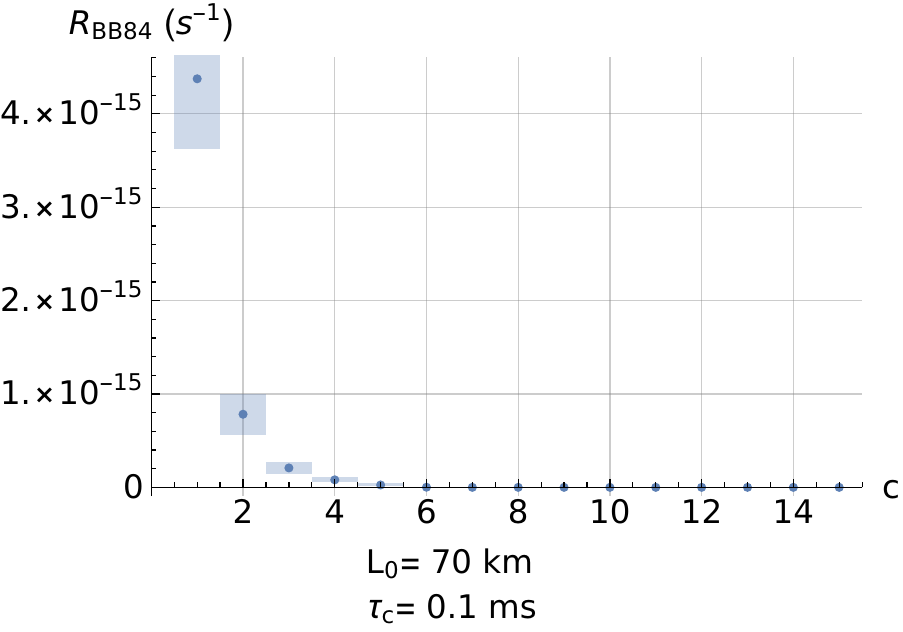}
	\includegraphics[width=\plotsize\textwidth]{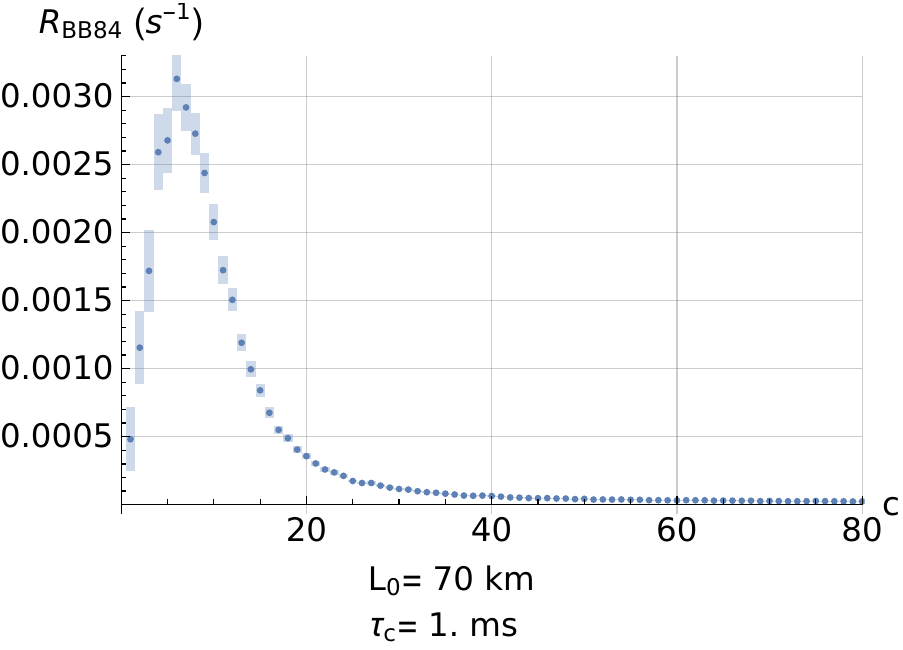}
}
\makebox[0.45\textwidth][c]{
\begin{minipage}{\baselineskip}
	\hfill\vspace{\baselineskip}
\end{minipage}
}
\makebox[0.45\textwidth][c]{
	\includegraphics[width=\plotsize\textwidth]{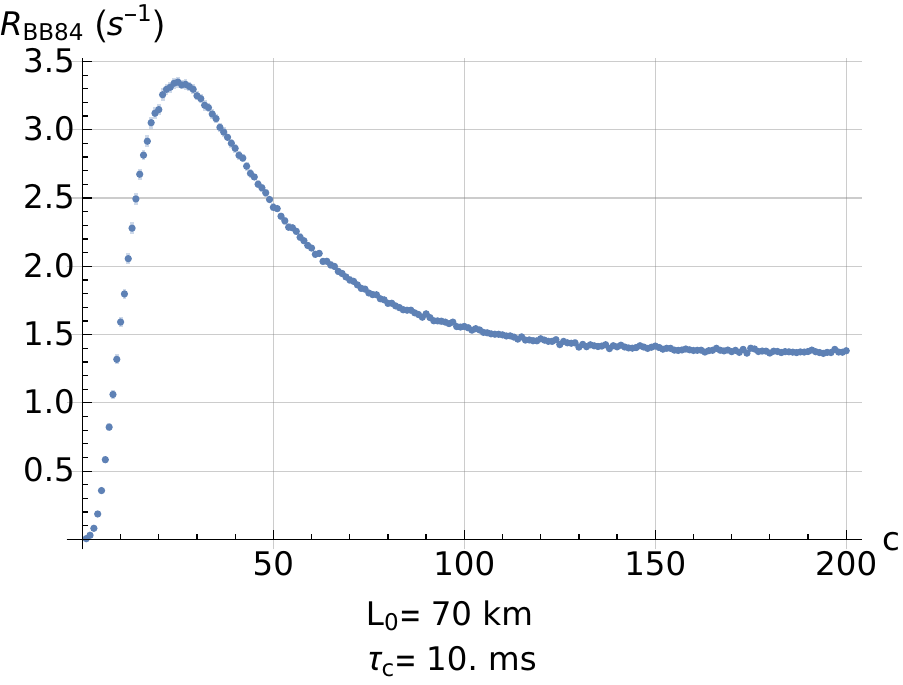}
	\includegraphics[width=\plotsize\textwidth]{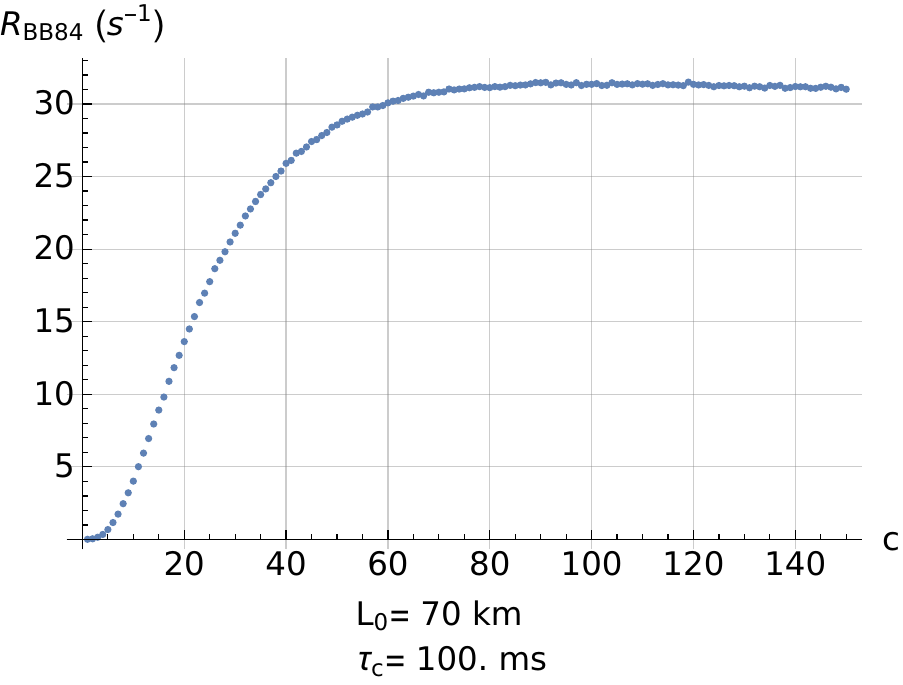}
}
	\caption{\label{fig:cutoff-70}Secret key rates of four-segment quantum repeaters in secret bits per second dependent on the memory cut-off parameter $c$ for segment length $L_0 = 70 \text{ km}$ and different coherence times of the memories for a bipartite quantum state $\tau_c$. Plotted with 3-$\sigma$ confidence intervals.}
\end{figure}

In this subsection, the secret key rates of quantum repeaters that are controlled by the cut-off policy are examined. In Figs. \ref{fig:cutoff-20} - \ref{fig:cutoff-70} the relation of the secret key rate and the cut-off parameter is displayed for various parameter choices of the segment length $L_0$ and the coherence time of the memories $\tau_c$.

In the limit of large cut-off parameters, the policy is identical to the case without cut-off. This is apparent in all plots, as the secret key rate converges to the no cut-off rates for large cut-off parameters.

For all segment lengths, a similar behavior with respect to the coherence time of the memories is visible. In the regime of low coherence times (i.e., $\tau_c = 0.1 \text{ ms}$), the secret key rates drop to values near zero, with the ideal cut-off parameter being the lowest possible value, which is one. This observation coincides with the results from Ref. \cite{Santra_2019}.  With increasing coherence time, the optimal cut-off parameter shifts towards larger values, and its peak of the secret key rate for the optimal cut-off decreases relative to the secret key rate of the asymptotically converging quantum repeater without cut-off. This decrease causes the peak to vanish for larger coherence times. At this point, the coherence time reaches a quality value beyond which the cut-off policy does not offer any improvements on the secret key rate.

This behavior moves towards longer coherence times as the segment distance increases. In other words, the shapes of the simulated and plotted secret key rates are qualitatively identical, interpreting the coherence time in relation to the segment length. This indicates that, conceptually, the results when conditioned on the relation between the parameters should be qualitatively applicable to any parameter regime.

\begin{figure}[tb]
\makebox[0.45\textwidth][c]{
	\includegraphics[width=0.264\textwidth]{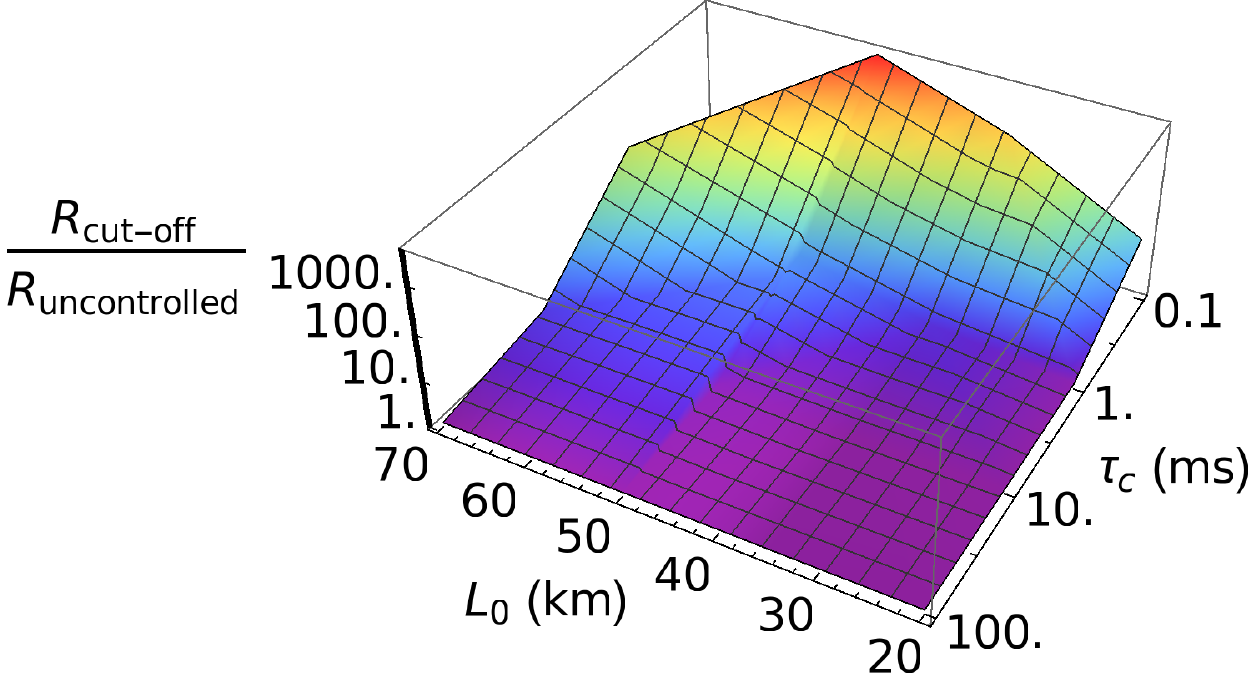}
	\includegraphics[width=0.176\textwidth]{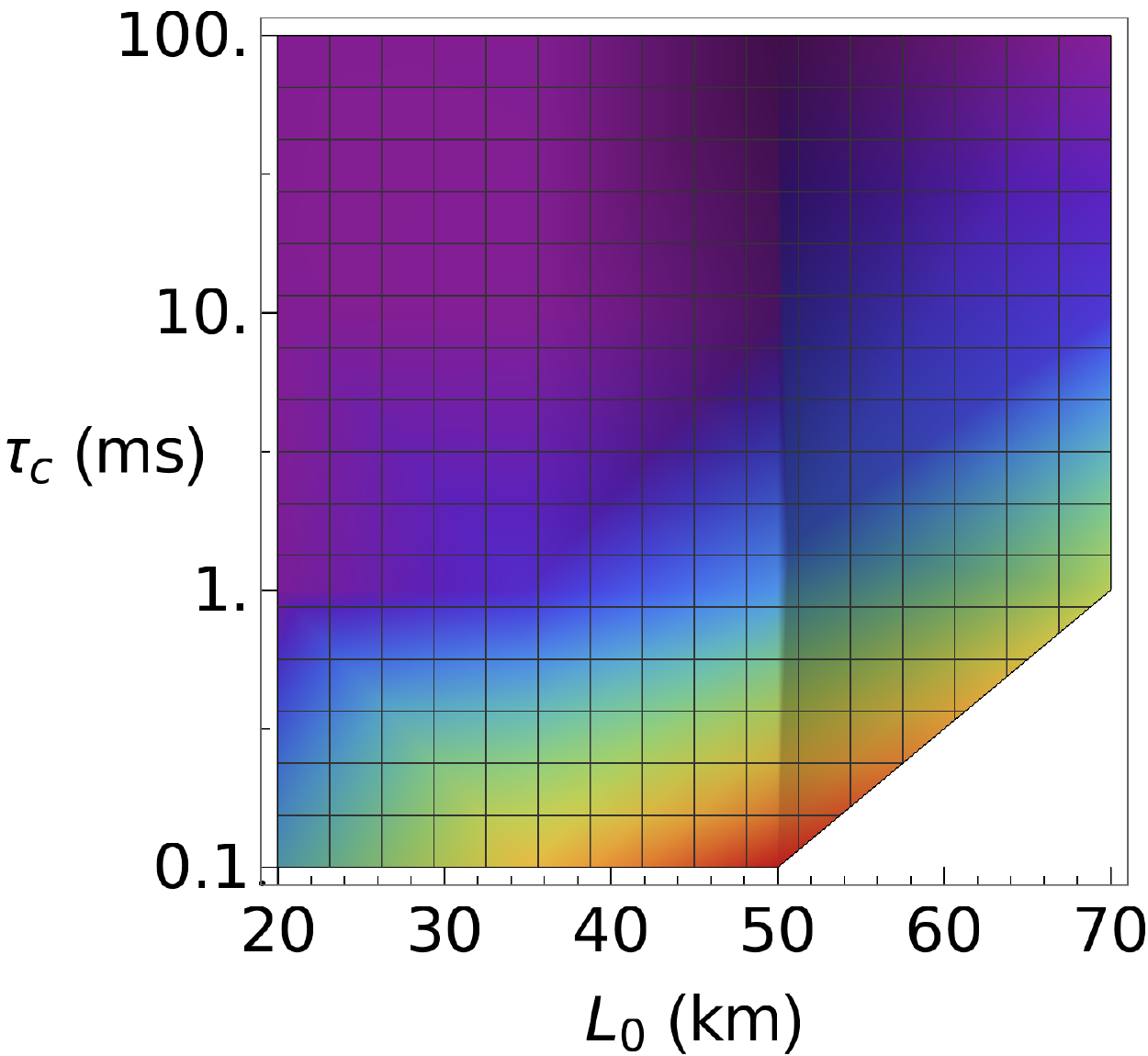}
	\hspace{30pt}
}
\makebox[0.45\textwidth][c]{
\begin{minipage}{\baselineskip}
	\hfill\vspace{\baselineskip}
	$R_{\text{cut-off}} \text{ / } R_{\text{no cut-off}}$
\end{minipage}
}
\makebox[0.45\textwidth][c]{
	\includegraphics[width=0.264\textwidth]{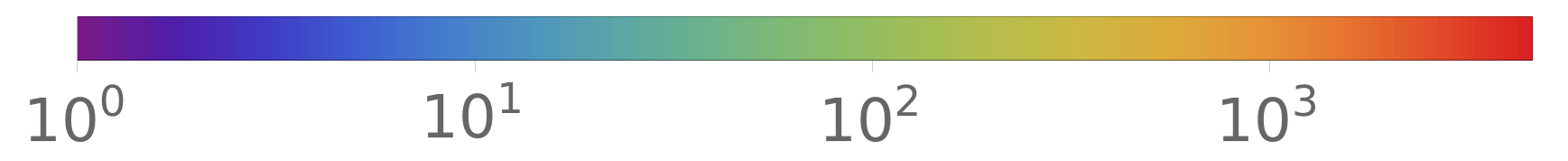}
}
	\caption{\label{fig:cutoffAdvantageLog}Simulated ratio between the secret key rates of the scheme with the best memory cut-off policy and schemes entirely without memory cut-off. This ratio and the coherence time of the memories for a bipartite quantum state $\tau_c$ are plotted logarithmically.}
\end{figure}

In Fig. \ref{fig:cutoffAdvantageLog}, the ratio between the secret key rate of the best cut-off policy and that without cut-off is shown. This further illustrates the above observations. The advantage of the cut-off vanishes for better quantum memories and shorter segment lengths and it increases for worse quantum memories and larger segment lengths. These observations coincide with the results from Ref. \cite{9495278}.

\subsection{Simulation - conclusion}
\label{sec:simulation-conclusion} 
In this section, the secret key rate of the BB84 protocol has been examined via numerical simulations for various parameters of a four-segment repeater setup. The two included imperfections were the dephasing of the quantum memories and the signal attenuation in the optical fibers.

The main observation that can be made with our results is that the worse the coherence time of the quantum memories is in relation to the segment length, the more significant is the advantage of the cut-off. This relation is steeper in the regime of bad memories in relation to the segment length and flattens for better memories until it converges to the point where the cut-off offers no improvement.

\section{Deep reinforcement learning applied to quantum repeaters}
\label{sec:deep-reinforcement-learning} 
In this section, we present our DRL approach to optimize secret key distribution employing quantum repeaters. The aim is a proof of concept, showing that DRL offers the possibility to provide sophisticated policies for memory treatment which outperform the more naive approaches.

First we will introduce our algorithm and its implementation. For the non-expert reader this includes a concise self-contained introduction to some DRL theory. After a brief discussion about the experimentation process the successful learning runs are presented. For conclusion, some observations about the learned policies are discussed.

\subsection{DRL algorithm}
\label{sec:deep-reinforcement-learning:alg}
\begin{figure}[tb]
	\fbox{\includegraphics[width=0.45\textwidth]{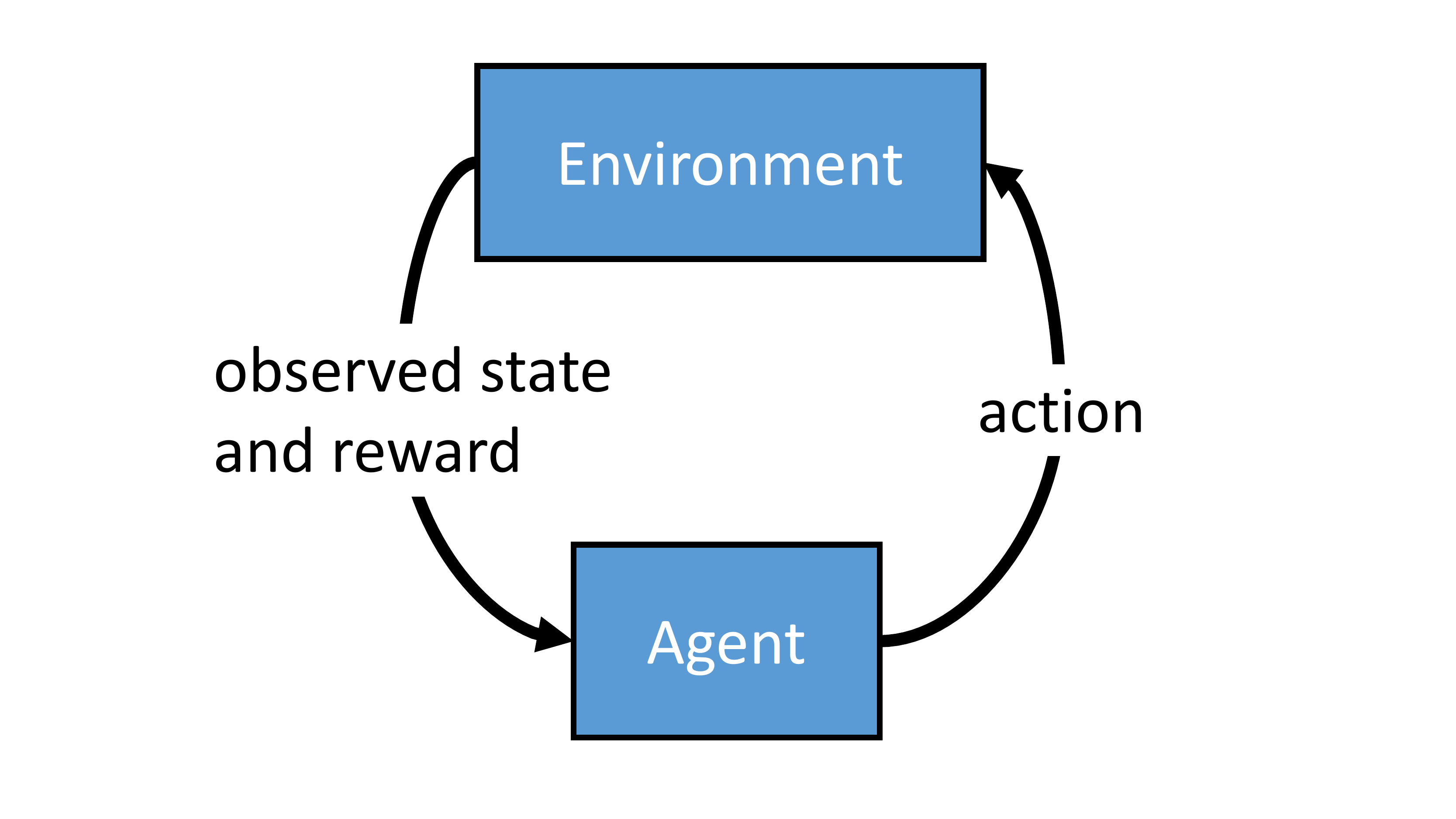}}
	\caption{\label{fig:theory-of-deep-reinforcement-learning:framework}Conceptual framework of an RL algorithm.}
\end{figure}

In this subsection, we will discuss the algorithm that we have employed. RL is a class of machine learning algorithms which trains a so-called agent to optimize its behavior in an environment. This concept is illustrated in Fig. \ref{fig:theory-of-deep-reinforcement-learning:framework}. Deep learning uses artificial neural networks to learn hierarchical representations in order to solve classification problems. In DRL these methods are combined by using an artificial neural network to model the agent \cite{10.1007/978-3-319-56991-8_32}. For our DRL application the MDP described in Sec. \ref{sec:mdp} serves as the model of the environment.

The rest of this subsection introduces some broadly known DRL theory with some intuitive explanations as well as the specific algorithm used in this work. For the interested reader Richard S. Sutton's and Andrew G. Barto's book \cite{Sutton1998} can serve as a comprehensive introduction to RL. A shorter, more practical introduction can be found on the website of Open Ai Spinning Up \cite{SpinningUp2018}.

We denote the policy of the agent as the probabilistic policy parameterized by a vector $\theta$,
\begin{equation}
	\pi_\theta(s,a) : S \times A \rightarrow [0,1] .
\end{equation}
The policy gives the probability of the agent taking the action $a \in A$ given the environment is in the state $s \in S$. In DRL the agent (i.e., the policy) is encoded in a neural network. In this case, the vector $\theta$ consists of the parameters (so-called weights) of the neural network.

Furthermore, we introduce the common definition of a trajectory as a sequence of states and actions up to a time $T$ called the (time) horizon,
\begin{equation}
	\tau = (s_0, a_0, s_1, a_1 .... ,s_T).
\end{equation}

The central merit of the optimization is the accumulated discounted reward of a trajectory $\tau$ up to a time $T$,
\begin{equation}
	R_0 = \left[ \sum^T_{t=0} \gamma^t r_t \right],
	\label{eq:R0}
\end{equation}
where $\gamma$ is the so-called discount parameter and $r_t$ the immediate reward at time $t$.

The usual objective of the agent is to optimize the policy $\pi_\theta$ to maximize the expected accumulated discounted reward over trajectories $\tau$,
\begin{equation}
	\underset{\theta}{\text{maximize}} \underset{\tau \sim \pi_\theta}{\mathds{E}} \left[ R_0 \right] ,
	\label{eq:DRL-objective}
\end{equation}
where $\underset{\tau \sim \pi_\theta}{\mathds{E}}$ expresses that the expectation value is taken over all trajectories $\tau$ following the policy $\pi_\theta$. Note that the trajectory implies all returned immediate rewards obtained in the trajectory. Further, note that the dependency on $\gamma$ and $T$ is left out of the expectation value, because it will be fixed in the optimization. These parameters which are fixed in the run time of the algorithm are termed hyperparameters in order to distinguish them from those parameters that the algorithm optimizes.

Policy gradient algorithms optimize the policy in terms of the objective via a gradient ascent,
\begin{equation}
	\theta_{k+1} = \theta_k + \alpha \nabla_\theta \underset{\tau \sim \pi_\theta}{\mathds{E}} \left[ R_0 \right]|_{\theta = \theta_k},
	\label{eq:policy-gradient}
\end{equation}
where $\alpha$ is a hyperparameter called the learning rate.

This already summarizes the concept of policy gradient optimization. The policy gradient $\nabla_\theta J(\pi_\theta)|_{\theta = \theta_k}$ is estimated on some training data, usually obtained from a simulation, and applied in an iterative fashion. One iteration is typically referred to as an epoch. As this gradient usually cannot be computed exactly and needs to be estimated, this introduces a finite variance into the stochastic gradient ascent. The fact that this gradient can be estimated is not obvious. The proof that this is in principle possible and the computations to do so are beyond what we want to cover in this section, and so we refer the interested reader to the aforementioned references \cite{Sutton1998,SpinningUp2018}.

Unfortunately, simple DRL algorithms often suffer from slow and unstable convergence properties, i.e. they tend to converge very slowly towards a local optimum and to overshoot in parameter updates causing them to often drastically lose progress in the learning process. In recent years novel, improved DRL algorithms have been proposed to counteract these issues \cite{8103164,schulman2017trust,schulman2017proximal}. Our algorithm of choice is a proximal policy optimization (PPO) \cite{schulman2017proximal}, which is a state of the art approach. PPO-clip is one of the proposed variants of Ref. \cite{schulman2017proximal} and is the one we use in this work. We will now explain the concept behind some of the improvements that were made in these algorithms.

A common approach is to generalize the objective of Eq. (\ref{eq:DRL-objective}):
\begin{equation}
	\underset{\theta}{\text{maximize}} \underset{\tau \sim \pi_\theta}{\mathds{E}} \left[ \Phi(\tau) \right],
\end{equation}
where we call $\Psi_t$ the (generalized) objective function. In general, any objective function $\Psi (\tau)$ satisfying
\begin{equation}
	 \nabla_\theta \underset{\tau \sim \pi_\theta}{\mathds{E}} \left[ R_0 \right]|_{\theta = \theta_k} =  \nabla_\theta \underset{\tau \sim \pi_\theta}{\mathds{E}} \left[ \Phi (\tau) \right]|_{\theta = \theta_k}.
	 \label{eq:gradient-equivalence}	
\end{equation}
is a valid objective function, since it leaves the gradient in Eq. (\ref{eq:DRL-objective}) invariant. Hence, one is free to choose this substitution, while still solving the same optimization problem.

In order to discuss how $\Phi(\tau)$ will be chosen we state the standard definitions of the on-policy value function $V^\pi(s_t)$, the on-policy action-value function $Q^\pi(s_t,a_t)$ and the advantage function $A^\pi(s_t,a_t)$.

We denote the accumulated reward after time $t$ as
\begin{equation}
	R_t = \sum^{T-t}_{l=0} \gamma^l r_{t+l}.
	\label{eq:Rt}
\end{equation}
Note that this is a generalization of Eq. (\ref{eq:R0}), containing $R_0$ as a specific case. This is already an improvement, since due to the truncation of the reward in Eq. (\ref{eq:Rt}), the evaluation of an action by the advantage function is independent of the trajectory prior to the action. This makes intuitively sense, because the evaluation of an action should only follow from what happened in consequence of that action.

The on-policy value function $V^\pi(s_t)$ is the expected reward received after time $t$ given that the environment at time $t$ is in the state $s_t$,
\begin{equation}
	V^\pi(s_t) = \underset{a_t,s_{t+1},...\sim \pi_\theta}{\mathds{E}} \left[ R_t \right].
\end{equation}
The on-policy action-value function $Q^\pi(s,a)$, is the expected reward received after time $t$ given the environment at time $t$ is in the state $s_t$ and the action $a_t$ is taken,
\begin{equation}
	Q^\pi(s_t,a_t) = \underset{s_{t+1},a_{t+1},... \sim \pi_\theta}{\mathds{E}} \left[ R_t \right].
\end{equation}

The difference between these functions defines the advantage function $A^\pi(s,a)$, which is a crucial merit in DRL,
\begin{equation}
	A^\pi(s_t,a_t) = Q^\pi(s_t,a_t) - V^\pi(s_t).
\end{equation}
Therefore, the advantage function evaluates how an action performs relative to the current policy. It can be shown that replacing the discounted reward $R_0$ in the objective in Eq. (\ref{eq:DRL-objective}) by the advantage function leaves the gradient in Eq. (\ref{eq:policy-gradient}) invariant \cite{schulman2015highdimensional} as demanded in Eq. (\ref{eq:gradient-equivalence}), i.e.
\begin{equation}
	 \nabla_\theta \underset{\tau \sim \pi_\theta}{\mathds{E}} \left[ R_0 \right]|_{\theta = \theta_k} =  \nabla_\theta \underset{\tau \sim \pi_\theta}{\mathds{E}} \left[ A^\pi (s_t, a_t) \right]|_{\theta = \theta_k}.
\end{equation}
To explain the appeal of this substitution we first note again that in most relevant applications the gradient in Eq. (\ref{eq:policy-gradient}) cannot be determined exactly. The solution is to perform a stochastic gradient ascent, i.e. for each parameter update ${\theta_k \rightarrow \theta_{k+1}}$ the gradient is estimated on a finite sample size. The benefit of using the advantage function in the objective is then to decrease the variance of these estimations, which results in better convergence properties.

This can be intuitively understood by interpreting the advantage function as a rescaling of the reward relative to the expected performance of the current policy. In the case where $Q^\pi(s_t,a_t)$, i.e. the performance merit conditioned on the action $a_t$, is better than the expectation value over all possible actions $V^\pi(s_t)$, the advantage function is positive, otherwise it is negative. It seems reasonable that we can think of the advantage function as a rescaled reward that becomes positive for any better-than-expected actions and negative for any worse-than-expected actions.

This is just one way in which DRL algorithms such as PPO can attempt to improve their convergence properties. More details can be found in the original PPO proposal \cite{schulman2017proximal}. We should also note that the advantage function is just one choice for the objective, as any function satisfying Eq. (\ref{eq:gradient-equivalence}) would be a valid candidate. A detailed discussion on objective functions can be found in Ref. \cite{schulman2015highdimensional}.

\subsection{Adaption of the DRL algorithm to non-additive rewards}
We will now discuss how our model of QKD via quantum repeaters can be applied to the DRL approach presented in the previous subsection.

A first approach could be to assign the fidelity of an entangled quantum state distributed between the communicating parties to the immediate reward in a time step and return zero reward if no entangled state was distributed in this time step. This approach was, for example, chosen in Ref. \cite{khatri2020policies} to optimize the distributed entanglement in a multiplexed point-to-point quantum communication channel.

In our case, we aim to optimize the secret key rate between the communicating parties. The distributed secret key, however, is not equivalent to the sum of the fidelities of the distributed entangled states. A simple solution consistent with the presented theory of DRL would be to consider a so-called episodic process where the reward of a trajectory is assigned to the last time step. Thus, one could assign the reward as
\begin{equation}
	r_t =
	\begin{cases}
	0 & t<T\\
	R_{\text{BB84}} \left( \tau \right) & t=T ,
	\end{cases}
\end{equation}

where $R_{\text{BB84}} \left( \tau \right)$ is the secret key rate of the distributed quantum states in the trajectory $\tau$. In Eq. (\ref{eq:detailed-skr}) it was shown how this can be calculated with the (accumulated)  storage time of the quantum states (or, equivalently, their fidelity) of the quantum states which corresponds to the rewards obtained in the trajectory $\tau$. In App. \ref{sec:markov-decision-process-modeling-a-quantum-repeater} it is explained how the rewards in a trajectory are determined by the simulated MDP.

The essence of the problem with this approach is often referred to as the credit assignment problem \cite{Minsky1961,Sutton1998}. Intuitively, this form of reward includes little information about the causal connection between a specific action and its influence on the reward. Mathematically this results in an increasing variance of the estimation of the gradient with longer time horizons.  Note that the key feature of the advantage function introduced in the preceding subsection is that it uses a time-grained evaluation of the process. Hence, an episodic reward severely hinders the means with which the convergence of the algorithm is improved. We were not able to achieve good convergence with this approach.

Therefore, in this work, we propose a novel approach where we incorporate a non-additive reward into the standard mathematical framework of RL. We redefine the optimization objective as
\begin{equation}
	\underset{\theta}{\text{maximize}} \underset{\tau \sim \pi_\theta}{\mathds{E}} \left[ R_{\text{BB84}} (\tau) \right].
\end{equation}
Therefore, the value functions now read
\begin{equation}
	V^\pi(s_t) = \underset{a_t,s_{t+1},...\sim \pi_\theta}{\mathds{E}} \left[ R_{\text{BB84}}(\tau_t) \right]
\end{equation}
and
\begin{equation}
	Q^\pi(s_t,a_t) = \underset{s_{t+1},a_{t+1},...\sim \pi_\theta}{\mathds{E}} \left[ R_{\text{BB84}}(\tau_t) \right],
\end{equation}
where $\tau_t = (s_{t+1},a_{t+1}, ... s_T)$ is the trajectory after time step $t$.

This new generalization replaces the accumulated reward after a time step by the secret key rate after that time step. Even though within the scope of this work we did not find a rigorous proof that this fulfills the condition of Eq. (\ref{eq:gradient-equivalence}) and thus is an equivalent optimization, it did deliver well converging results empirically. Based on this approach we achieved reasonable convergence with good results which we will present in Sec. \ref{sec:drl-results}. It is important to note that the lack of a rigorous proof for the optimization does not weaken the numerical results obtained, since the performance of a policy is evaluated independent of how the policy is found.

In our implementation, as is common practice, we used an actor-critic function $A_t$ to estimate the advantage function,
\begin{equation}
	A_t = R_{\text{BB84}}(\tau_t) - V^\pi(s_t),
	\label{eq:At}
\end{equation}
where an artificial neural network is employed to approximate $V^\pi(s_t)$, which is trained on the same training data the agent is trained on. More details on the implications and properties of our advantage function and on the topic of discounting rewards can be found in the App. \ref{sec:A:discounting}. Below the Box Algorithm \ref{alg:PPO} shows the pseudo-code of our implementation. Finally, we would like to stress that generally this adaption could be applied to other optimization problems with non-additive rewards.

\begin{figure*}
\begin{algorithm}[H]
	\While{not converged}{
		\begin{enumerate}
			\item Collect set of trajectories $D_k = \{ \tau_i \}$ by running policy $\pi_k = \pi \left(\theta_k \right)$ in the environment.
			\item Compute reward functions $R_t = R_{\text{BB84}} \left( \tau |_{t^\prime \geq t} \right)$ for every trajectory $\tau \in D_k$ and time step $t \in \left[ 0, T \right]$.
			\item Compute $A_t = R_t-V_{\phi_k}\left( s_t \right)$.
			\item Update the policy by maximizing the RL objective
			\begin{equation}
				\theta_{k+1} = \underset{\theta} {\text{argmax}} \frac{1}{|D_k| T} \sum_{\tau \in D_k} \sum_{t=0}^T \text{min} \left( \frac{\pi_\theta \left( a_t | s_t \right)}{\pi_{\theta_k} \left( a_t | s_t \right)} A_t , \text{clip} \left( \frac{\pi_\theta \left( a_t | s_t \right)}{\pi_{\theta_k} \left( a_t | s_t \right)} , 1-\epsilon , 1+\epsilon \right) A_t  \right)
				\label{eq:PPO-alg}
			\end{equation}
			via Adam stochastic gradient ascent \cite{kingma2014adam}.
			\item Fit value function by regression on mean-squared error,
			\begin{equation}
				\phi_{k+1} = \underset{\phi} {\text{argmin}} \frac{1}{|D_k| T} \sum_{\tau \in D_k} \sum_{t=0}^T \left( V_\phi \left( s_t \right) - \Phi_t \right) ^2
			\end{equation}
			via Adam stochastic gradient descent \cite{kingma2014adam}.
		\end{enumerate}
	}
	\caption{\label{alg:PPO}PPO Algorithm adapted from Ref. \cite{SpinningUp2018}}
\end{algorithm}
\end{figure*}

\subsection{Application of the algorithm to our physical model}
In every time step the agent is given the decision to discard any of the bipartite quantum states stored in the quantum repeater based on the observation of the current state of the environment. It should be pointed out that this allows the agent to develop a vast range of complex policies. The agent is able to make decisions about each quantum state based on an arbitrary selection of the available information. \footnote{These policies can even emulate other repeater schemes. For example, a sequential repeater scheme like that in Refs. \cite{PhysRevA.102.042614,https://doi.org/10.48550/arxiv.2203.10318}. This case is exactly emulated by discarding any quantum states that are generated in segments where the sequential  scheme would not attempt any entanglement generation.} The decision about when to perform a swapping operation is not under control of the agent. As discussed in Sec. \ref{sec:dephasing-strategies}, the chosen swapping strategy is fixed to "swap as soon as possible".

For the present work, we decided to give the agent at any time the entire information about the current state of the environment. This means that the agent can make decisions on how to operate a quantum memory based on information that physically would not be immediately available at that time for a local operation on the memory. This assumption can be motivated by two primary reasons. The first is that the all-knowing agent can more easily find good solutions. When trying to implement algorithms that are sensitive to their tuned hyperparameters, it is usually a good approach to start with simpler problems and increase the complexity when the first results are obtained. The second reason is that one can possibly learn more about the behavior of the problem within the parameter space when as much information as possible is given into the analysis. This may also be considered as a first step towards methods where an agent is trained to mimic the behavior of the all-knowing agent as well as possible while having only the restricted information of a realistic scenario available. This two-step training of an agent could prove more efficient than trying to learn directly on the realistically restricted information. Lastly, it would be highly non-trivial to also include all classical communication in the simulation.

The benchmarks for the agents to surpass are the best naive strategies as determined in Sec. \ref{sec:simulations:results}, which includes schemes both with and without cut-off.

\textit{Remark:} It is not a priori known for which experimental setup the cut-off policy does not provide any benefit. Therefore, there is no reasonable measure for which the no-cut-off approach is as good as the cut-off approach. The cut-off policy can never become worse, as it emulates the no-cut-off approach for large cut-off parameter. Thus, we chose to take the maximum of all data points simulated in Sec. \ref{sec:simulating-key-distribution-based-on-quantum-repeaters} for each experimental setup as the benchmark. This includes cut-off and no-cut-off strategies. For those cases where the cut-off does not offer an improvement, this means that the approximately optimal cut-off was run multiple times as the secret key rate converged for larger cut-offs, as discussed in Sec. \ref{sec:simulations:results}. Since from these runs the maximum and not the average was taken, the benchmark is slightly biased towards better secret key rates for these cases. We expect this bias to be negligible for the comparison, since the uncertainties are small enough to prevent significantly overestimated rates.

\subsection{Implementation and experimentation}
\subsubsection{Implementation}
The learning algorithm of this work is a python implementation within the open-source DRL framework "OpenAI Spinning Up" \cite{SpinningUp2018} by "Open AI" which uses the "Open AI Gym" \cite{https://doi.org/10.48550/arxiv.1606.01540}. The Spinning Up framework provides some standard learning algorithms, a logger to store a variety of diagnostics during a learning run, as well as plotting tools to display them. Our implementation is a modified version of the PPO implementation provided by Spinning Up.

Internally the neural network and the sampling of actions is implemented using the "TensorFlow" \cite{tensorflow2015-whitepaper} library.

The features that had to be modified and added to the implementation of the Spinning Up software will be explained in the following:

\begin{itemize}
\item The calculation of the secret key rate, raw rate, and average fidelity of a trajectory was implemented and included in the main loop of experience collection and storage.
\item The neural network implemented in the framework did not support multidimensional action spaces. Therefore, the architecture of the neural network and the calculation of the action probabilities $\pi_\theta$ were generalized and adjusted.
\item The PPO algorithm used a generalized advantage estimation \cite{schulman2015highdimensional}, which is incompatible with the non-additive rewards of QKD. Hence, this was replaced by our generalized actor-critic function of Eq. (\ref{eq:At}).
\end{itemize}

The experience collection loop and the stochastic gradient ascent via Adam support multi-process parallelization via MPI (message passing interface) \cite{10.1145/169627.169855}.

The final evaluation of the performance of a learned policy was done by loading its neural network into the simulation framework of Sec. \ref{sec:simulating-key-distribution-based-on-quantum-repeaters}. Each agent was evaluated over 100 trajectories. The length of the trajectories for the simulation of each agent's policy can be found in App. \ref{sec:A:sim-trajectory-lengths}. Moreover, a program was implemented to store all the states seen by an agent and the count of actions it took separated for each state. This will be later used to analyze the behavior of the agents.

\textit{Remark:} The implementation also stores the neural network representing the agent at intermediate epochs. Thus, one could analyze the agent at predetermined points in the learning process.

\subsubsection{Hyperparameters}
Below we list the tuned hyperparameters of the implemented algorithm. For details of those parameters which are not discussed in this section, we refer the reader to the original proposal of the algorithm \cite{schulman2017proximal} and the documentation of the original implementation \cite{SpinningUp2018}:
\begin{itemize}
\item Number of hidden layers, as well as the number of neurons for each hidden layer for the neural networks of the agent and value function estimation. Note that the numbers of neurons in the input and output layer are fixed by the state- and action-space of the environment, respectively.
\item Activation functions for each layer of the two neural networks.
\item Length of the simulated trajectories.
\item Number of simulated trajectories per epoch.
\item Learning rate for the policy.
\item Learning rate for the value function $V^\pi(s_t)$.
\item $\epsilon_{\text{clip}}$: Hyperparameter of the clipped objective of PPO in Eq. (\ref{eq:PPO-alg}).
\item Number of gradient steps in the policy update in one epoch.
\item Number of gradient steps in the value function update in one epoch.
\item Maximum Kullback-Leibler divergence which stops the gradient update steps of the policy early if exceeded.
\item $\epsilon_{\text{Adam}}$: A parameter in the Adam optimization to improve numerical stability.
\end{itemize}

The agent as well as the value function estimation are represented by a four-layer (output and input layer, two hidden layers) artificial neural network. The hidden layers consist of 32 neurons each. For the activation function of the hidden layers of the agent the tangens hyperbolicus was chosen. The output layer of the agent uses a sigmoid activation in order to interpret the outputs as probabilities. These hyperparameters yielded good convergence properties empirically. The dimension of the input and output layers of the agent corresponds to the dimensions of state and action space of the MDP, respectively. This is 10 and 9 for a four-segment quantum repeater, respectively. The input layer of the value function estimation is also determined by the dimension of the MDP state space and the output dimension is one, thus yielding a scalar value of the value function. The choices for the rest of the hyperparameters are listed in Sec. \ref{sec:A:drl-hyperparameter}. 

\subsubsection{Experimentation}
The process of the simulated quantum repeaters is highly probabilistic, resulting in high variances for the results of simulated trajectories of short length. This, together with the lack of discounting, leads to slow convergence of the algorithm.

In consequence, hyperparameter tuning turned out to be an extensive task in the setting of this work. The final tuning achieved visible learning progress in the order of an hour, which gives an impression of how time-consuming and challenging examining the vast hyperparameter space is where learning progress is significantly less.

With computation time being a major limiting factor and the algorithm being inherently well suited for parallelization, the use of a high-performance computation cluster could very well enable significantly more efficient experimentation. Unfortunately, the implementation on a cluster was not possible for this work.

The used TensorFlow implementation of the Adam optimizer was not always numerically stable. That caused the neural network to have "NaN" entries, forcing the learning algorithm to stop. This was counteracted by empirically adjusting the hyperparameter $\epsilon_{\text{Adam}}$, which helped but did not solve the problem entirely. As the functionality to resume aborted learning processes was not implemented within the scope of this work, the occurrence of this case was the reason that some of the learning processes presented in this section were not continued, even though they did not converge.

\subsection{Results}
\label{sec:drl-results}
The DRL implementation was applied to some of the parameter points of Sec. \ref{sec:simulations:results}. The neural network of all stored agents of the learning runs as well as the gym environment simulating the quantum repeaters are publicly available in Ref. \cite{myGitRep}.

\subsubsection{Performance and learning progress of the agents}
\begin{figure}[tb]
\makebox[0.4\textwidth][c]{
	\begin{subfigure}[b]{0.25\textwidth}
		\caption{\label{fig:4-20-100-drl}}
		\includegraphics[width=\textwidth]{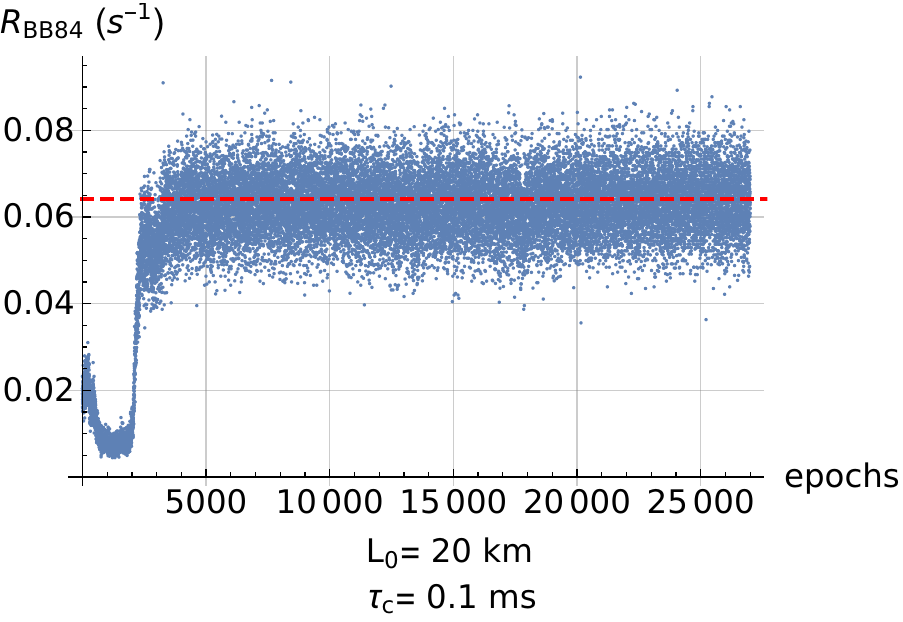}
	\end{subfigure}
	\begin{subfigure}[b]{0.25\textwidth}
		\caption{\label{fig:4-20-10000-drl}}
		\includegraphics[width=\textwidth]{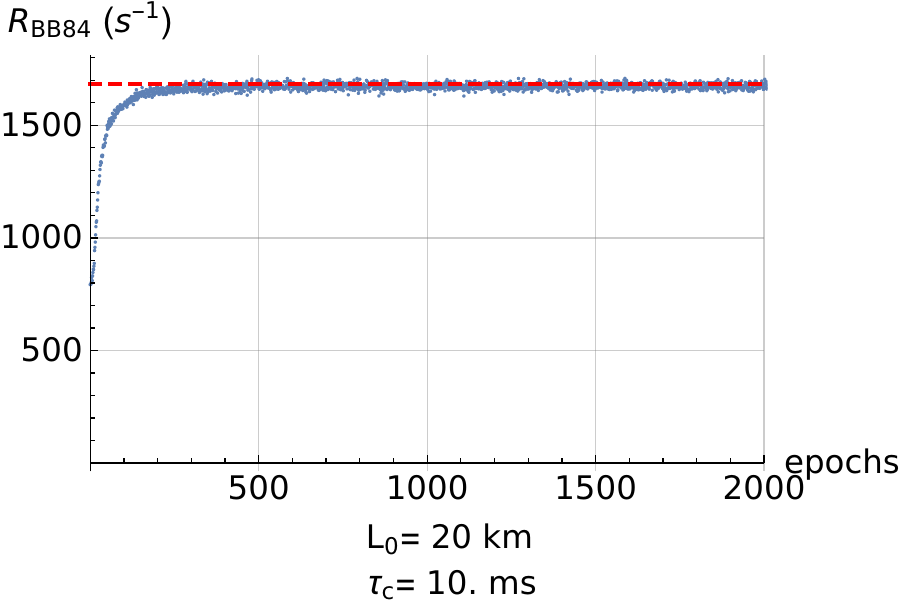}
	\end{subfigure}	
}
\makebox[0.4\textwidth][c]{
	\begin{subfigure}[b]{0.25\textwidth}
		\caption{\label{fig:4-35-10000-drl}}
		\includegraphics[width=\textwidth]{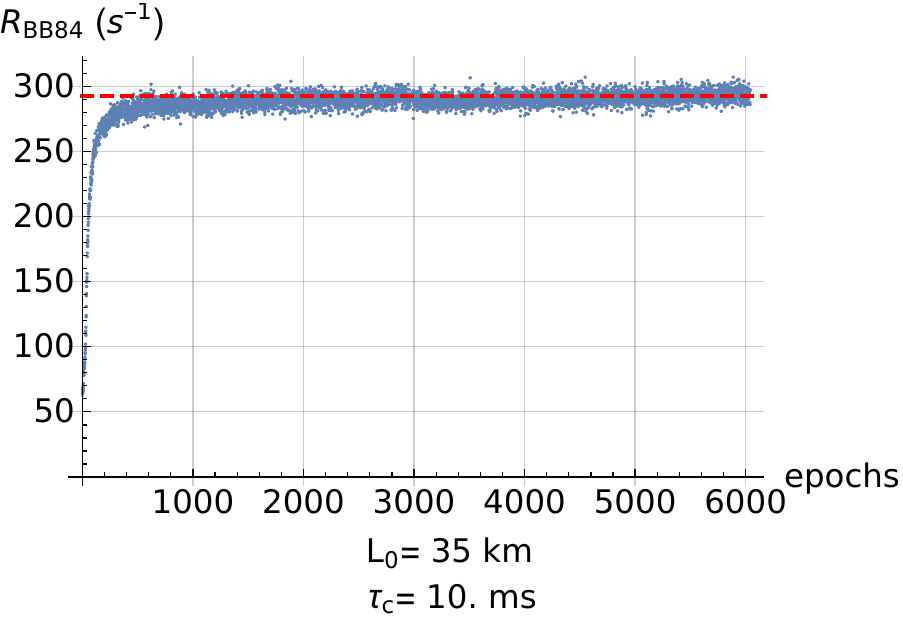}
	\end{subfigure}
	\begin{subfigure}[b]{0.25\textwidth}
		\caption{\label{fig:4-50-100000-drl}}
		\includegraphics[width=\textwidth]{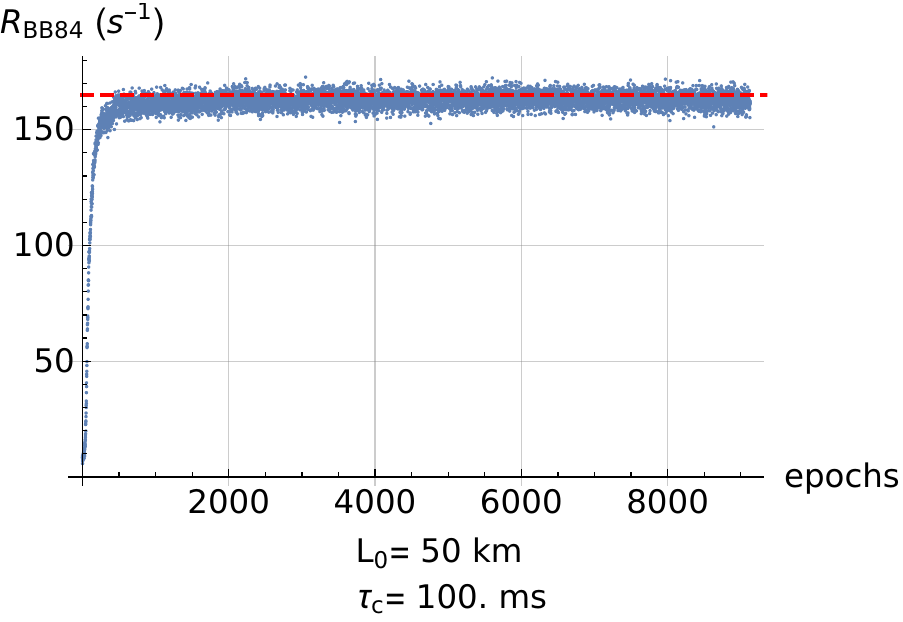}
	\end{subfigure}	
}
	\caption{\label{fig:learning-reproduced}Learning progress where the performance of the learned policies approximately matched the benchmark marked by the red dashed line. The average over the secret key rate of all trajectories for each epoch is plotted. An epoch is one iteration of the main loop of the algorithm. The runs \ref{fig:4-20-100-drl}, \ref{fig:4-20-10000-drl} and \ref{fig:4-50-100000-drl} converged without breaking numerically. For these, a long stagnating part after the displayed epochs is not shown. $\tau_c$ is the coherence time of the memories for a bipartite quantum state.}
\end{figure}

\begin{figure}[tb]
\makebox[0.4\textwidth][c]{
	\begin{subfigure}[b]{0.25\textwidth}
		\caption{\label{fig:4-20-1000-drl}}
		\includegraphics[width=\textwidth]{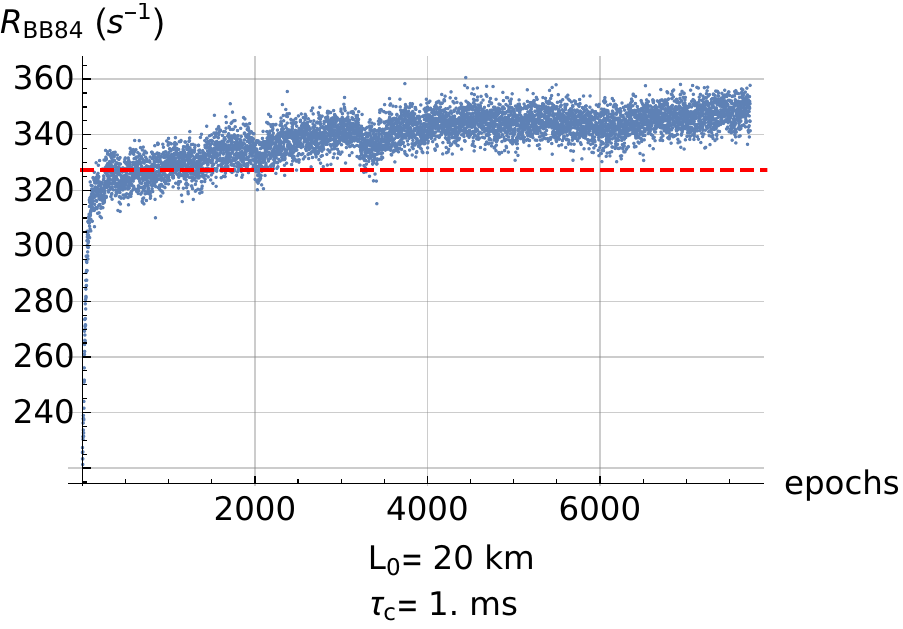}
	\end{subfigure}
	\begin{subfigure}[b]{0.25\textwidth}
		\caption{\label{fig:4-35-1000-drl}}
		\includegraphics[width=\textwidth]{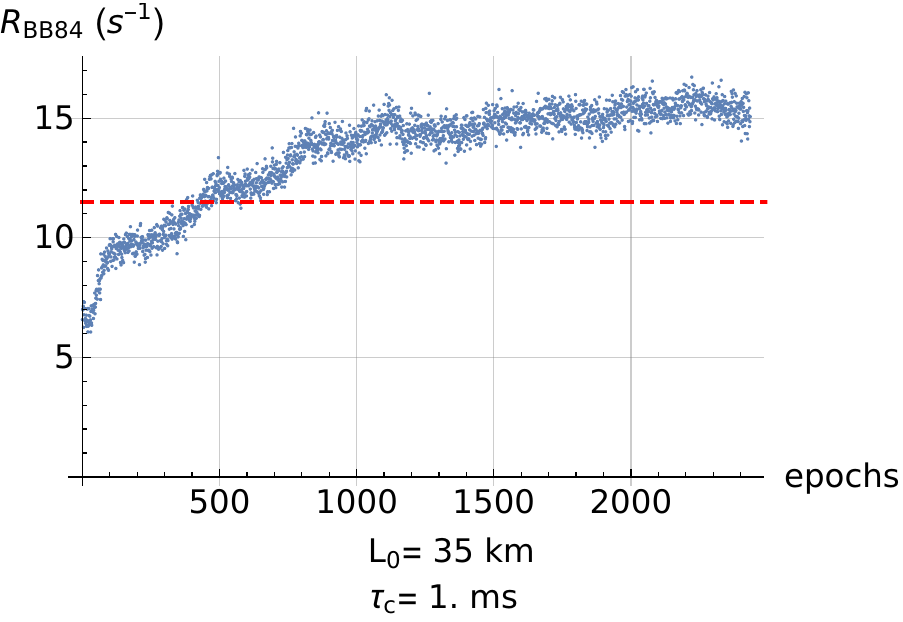}
	\end{subfigure}	
}
\makebox[0.4\textwidth][c]{
	\begin{subfigure}[b]{0.25\textwidth}
		\caption{\label{fig:4-50-1000-drl}}
		\includegraphics[width=\textwidth]{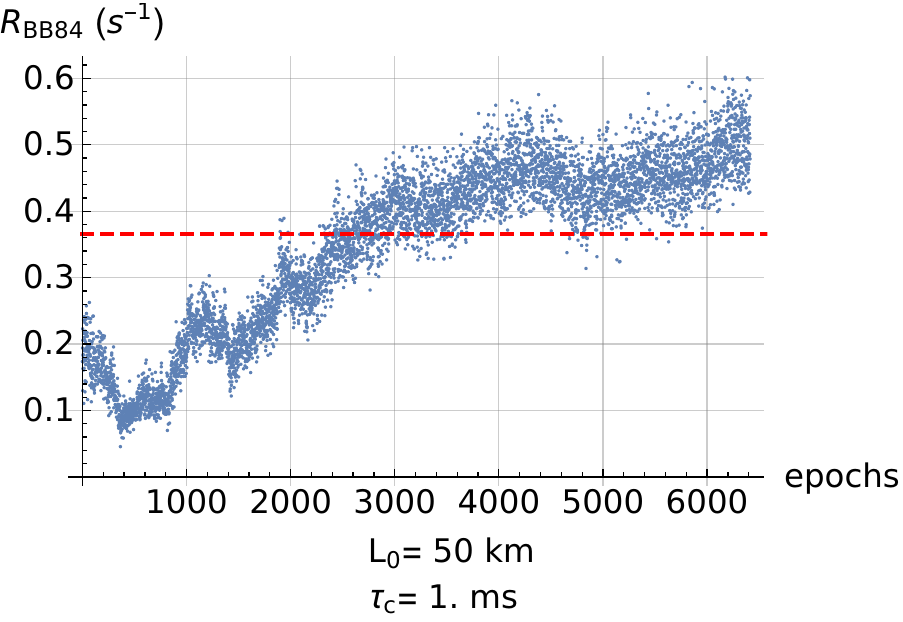}
	\end{subfigure}	
		\begin{subfigure}[b]{0.25\textwidth}
		\caption{\label{fig:4-50-10000-drl}}
		\includegraphics[width=\textwidth]{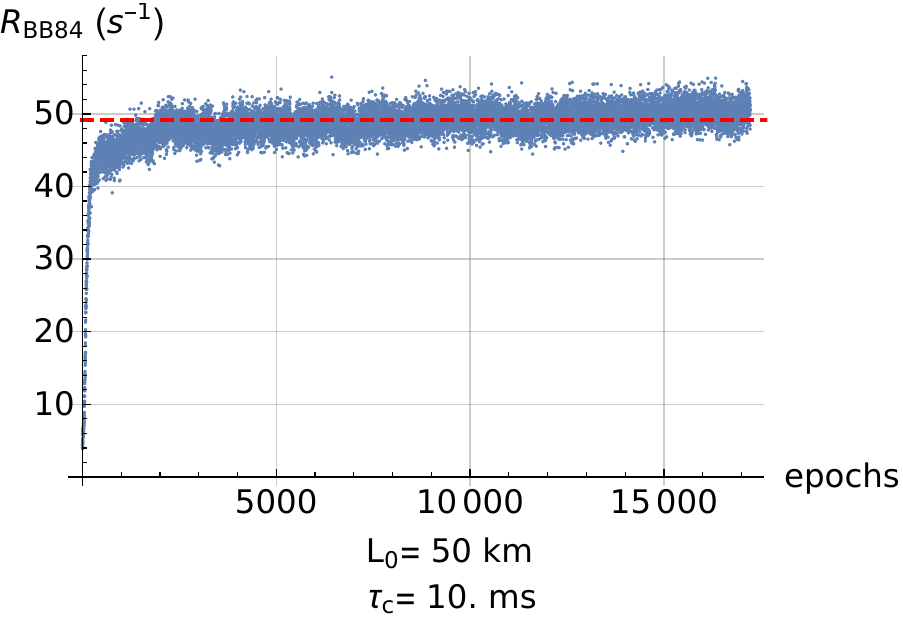}
	\end{subfigure}
}
\makebox[0.4\textwidth][c]{
	\begin{subfigure}[b]{0.25\textwidth}
		\caption{\label{fig:4-50-1452-drl}}
		\includegraphics[width=\textwidth]{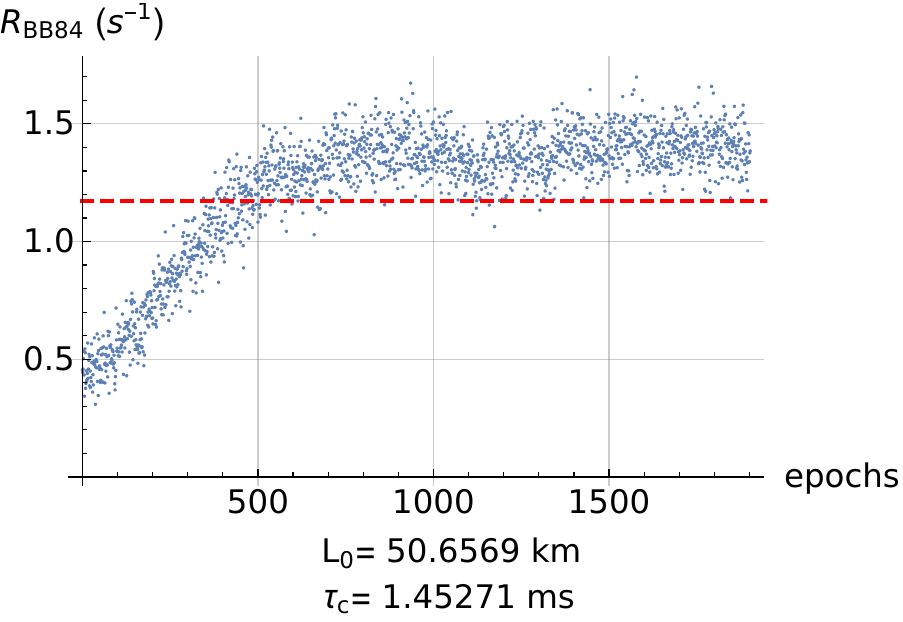}
	\end{subfigure}	
}
	\caption{\label{fig:learning-surpassed}Learning progress where the performance of the learned policies surpassed the benchmark marked by the red dashed line. The average over the secret key rate of all trajectories for each epoch is plotted. An epoch is one iteration of the main loop of the algorithm. $\tau_c$ is the coherence time of the memories for a bipartite quantum state.}
\end{figure}

The learned policies can be classified into two categories. In the limiting cases where the naive cut-off offers no improvement on the secret key rate or where the optimal cut-off is one, the learning algorithm approximately reproduced these results. The learning process of these runs is plotted in Fig. \ref{fig:learning-reproduced}. In other cases where the naive cut-off offered a significant improvement, the learning algorithm found a policy with even better performance. These cases are plotted in Fig. \ref{fig:learning-surpassed}. The list of hyperparameter choices for each run, as well as some discussion on the runtime,  can be found in App. \ref{sec:A:drl-hyperparameter}.

Tables \ref{tab:learning-reproduced} and \ref{tab:learning-surpassed} display the comparison of the learned policies with the benchmark strategies and a no-cut-off strategy.

\begin{table}[tb]
\begin{tabular}{lcccc}
	\toprule
	Figure & $L_0$ (km) & $\tau_c$ (ms) & $\frac{R_{\text{agent}}}{R_{\text{no cut-off}}}$ & $\frac{R_{\text{agent}}}{R_{\text{benchamrk}}}$\\
	\midrule
	\ref{fig:4-20-100-drl} & $20$ & $0.1$ & $18.0 \pm 0.2$ & $0.984 \pm 0.006$\\ \addlinespace
	\ref{fig:4-20-10000-drl} & $20$ & $10$ & $1.001 \pm 0.002$ & $0.997 \pm 0.002$\\ \addlinespace
	\ref{fig:4-35-10000-drl} & $35$ & $10$ & $1.014 \pm 0.004$ & $1.001 \pm 0.004$\\ \addlinespace
	\ref{fig:4-50-100000-drl} & $50$ & $100$ & $0.996 \pm 0.003$ & $0.984 \pm 0.003$\\
	\bottomrule
	\sbline
\end{tabular}
	\caption{\label{tab:learning-reproduced} Comparison between the learned policies and the benchmarks for those cases where the benchmark was approximately matched. $\tau_c$ is the coherence time of the memories for a bipartite quantum state.}
\end{table}

\begin{table*}[tb]
\begin{tabular}{lcccccc}
	\toprule
	Figure & $L_0$ (km) & $\tau_c$ (ms) & $R_{\text{agent}} \left( s^{-1} \right)$ & $\frac{R_{\text{agent}}}{R_{\text{no cut-off}}}$ &$r = \frac{R_{\text{agent}}}{R_{\text{benchamrk}}}$ & $\frac{r-1}{\Delta r}$\\
	\midrule
	\ref{fig:4-20-1000-drl} & $20$ & $1$ & $350 \pm 8$ & $1.172 \pm 0.004$ & $1.068 \pm 0.003$ & $20.4453$\\ \addlinespace
	\ref{fig:4-35-1000-drl} & $35$ & $1$ & $15.1 \pm 0.7$ & $2.67 \pm 0.03$ & $1.31 \pm 0.01$ & $32.2178$\\ \addlinespace
	\ref{fig:4-50-1000-drl} & $50$ & $1$ & $0.51 \pm 0.01$ & $12.3 \pm 0.1$ & $1.382 \pm 0.009$ & $42.9056$\\ \addlinespace
	\ref{fig:4-50-10000-drl} & $50$ & $10$ & $50.3 \pm 0.8$ & $1.180 \pm 0.003$ & $1.024 \pm 0.002$ & $10.4957$\\ \addlinespace
	\ref{fig:4-50-1452-drl} & $52.6569$ & $1.45271$ & $1.43 \pm 0.04$ & $6.45 \pm 0.04$ & $1.218 \pm 0.005$ & $40.0445$\\
	\bottomrule
	\sbline
\end{tabular}
	\caption{\label{tab:learning-surpassed}Comparison between the learned policies and the benchmarks in the cases where the benchmark was surpassed. $\tau_c$ is the coherence time of the memories for a bipartite quantum state.}
\end{table*}

Table \ref{tab:learning-reproduced} shows for the first category that the performance of the learned policies is approximately the same as the benchmark. The cases where the benchmarks were matched are exclusively those cases where the parameters correspond to limiting cases identified in Sec. \ref{sec:simulations:results}, where the optimal strategy is either having no cut-off or the minimal possible cut-off. Therefore, it seems reasonable that in these regimes, there is no more room for improvement for the policies.

Table \ref{tab:learning-surpassed} presents the advantage of the learned policies that surpassed the benchmark. The narrow confidence intervals of the ratios indicate that the learned policies outperform the benchmarks. This is further shown via the quantity $\frac{r-1}{\Delta r}$, which is the difference between the measured ratio and the ratio a strategy with no benefit over the benchmark would yield in units of its uncertainty. This, therefore, prevents the conclusion that the result is a statistical fluctuation. The improvements over the benchmarks range from 2\% to 38\%. Furthermore, apparently the more improvement the naive cut-off offers over a quantum repeater without cut-off, the more potential lies in the learned policies for this quantum repeater, since this correlation is visible in the achieved and presented examples. It is worth mentioning that this is an analog result to what was stated in Sec. \ref{sec:simulation-conclusion} for the naive cut-off policy.

\subsubsection{Discussion on the learned policies}
Extracting an in-depth understanding of a policy from the neural network of the agent is a non-trivial task and an open question of current research \cite{dawid2020phase}. Nevertheless, some aspects gained by looking at the policies of the presented results shall be discussed in this subsection.

The DRL algorithm optimizes probabilistic policies, where the probability can be interpreted as how certain an agent is about the optimality of an action for a given state. In order to output probabilities, a sigmoid function is applied to the output layer of the neural network representing the policy. The sigmoid function converges asymptotically to one and zero for large absolute output weights. First, this means that in order to get an approximately guaranteed probability, comparably large parameter changes are necessary. Secondly, a real guaranteed probability can, in theory, never be achieved. This causes even well converged policies to take actions that they evaluate to be non-ideal occasionally. This is useful in the learning process, as the agent occasionally reviews these actions, but slightly reduces the final performance of the agent. A suggestion to improve this is to map probabilities above a high threshold to one and analogously for low probabilities to zero in the final policies. Nonetheless, this has not been done in this work.

The actions of those policies that matched the benchmark were confirmed to reproduce their respective benchmark strategy approximately. The agents that surpassed the benchmark are more challenging to comprehend. A complete understanding of the policies could not be achieved in this work. Nevertheless, three observed patterns that give insights about how the policies achieved better performance are given in the following:

\begin{itemize}
\item The policies keep quantum states with higher storage times shared between nodes that are farther apart compared to quantum states shared between closer nodes, which are discarded much earlier.
\item The policy's decision to keep or discard appears to be influenced by the existence and quality of other quantum states in the repeater. If, for example, in the proximity of a quantum state, another quantum state exists, the agent is more likely to keep the quantum state.
\item The policy discards a quantum state comparably early if a single bipartite quantum state is shared between the two nodes surrounding the first quantum state.
\end{itemize}

These strategies might seem intuitive or even obvious and could probably be thought of without a machine learning approach. The advantage of the machine learning approach, however, is the scaling of these strategies. Even though the concept might be clear, finding exact numbers remains a task on its own, for which DRL is used here.

On the other hand, some decisions of the agents were obviously unreasonable. This includes decisions of well converged probabilities. It shows that the full potential of the DRL approach has not been achieved yet in this work and that there is still room for improvement in the algorithm and policies. The following examples are two decisions that an agent made with high probability that fall into this category:

\begin{itemize}
\item In the case mentioned above, where a policy would discard a quantum state $A$ if it is the only one left in the repeater but then keeps it because of another quantum state $B$ in the proximity, some unreasonable actions occurred. In some cases, it kept the state $A$ but discarded the state $B$, even though the decision to keep state $A$ was dependent on the existence of state $B$.
\item In a scenario where two bipartite states with exactly symmetric positions in the repeater where stored, the agent decided to discard the one with the lower and to keep the one with the higher storage time.
\end{itemize}

Another observation worth mentioning is that the policies tended to be asymmetric in the sense that two symmetric segments were often treated differently. This might happen because of non-ideal convergence, but one should note that nothing contradicts the possibility that an asymmetric policy might be optimal, despite the structure of a four-segment repeater being symmetric.

Furthermore, the policies were less converging in their probabilities than their convergence in performance might suggest. This could be explained by the fact that two different actions do not necessarily cause different secret key rates. Some actions might be equivalent or very close in performance. Thus, the agent will not, or will only very slowly, learn to choose one action over the other.

\subsection{Future work}
One suggestion for further improvements is based on the fact that states of the environment have significantly different frequencies in a trajectory. Therefore, the agent has superfluous experience about prevalent states of the MDP while having very sparse information about others. This creates an imbalance and thus sample inefficiency. A suggestion to counteract this is to create an environment that uses a random initial state of the trajectory in ways that flatten the distribution of the states the agent sees.

The next step towards analyzing more realistic repeater chains would be to include classical communication by having multiple cooperating agents where each agent has access only to the information available at one corresponding repeater node.

\subsection{Deep reinforcement learning - conclusion}
A proof of concept that DRL can be exploited for QKD via quantum repeaters to find sophisticated policies that improve the secret key rate over "naive" approaches has been achieved. The results of the learned policies ranged from a reproduction of performance up to a 38\% improvement over the naive simulation approach of Sec. \ref{sec:simulating-key-distribution-based-on-quantum-repeaters}. Furthermore, we have identified shortcomings of the policies and suggested enhancements to the algorithm to further optimize policies, showing potential room for improvement. Additionally, utilizing a high-performance computation cluster would probably go a long way in enhancing the effectiveness with which solutions can be found. This provides motivation to further pursue DRL algorithms in control problems of multi-segment quantum repeaters.

The improvements of the learned policies as presented here might seem small for making a strong impact on long-distance QKD applications. This, however, could be misleading, as the examined repeaters only had four segments. Practical applications most likely will use considerably more segments. This increases the number of possible policies and, in particular, the asymmetry between the repeater nodes. Therefore, we would intuitively expect the improvements to increase for more segments, as the parameter space grows in complexity. Additionally, the complexity of quantum repeaters generally grows with the number of segments, making other approaches less feasible, while DRL is especially suited for problems that can be simulated but no longer analytically analyzed. This emphasizes the meaning of a proof of concept. It has been conceptually shown that DRL can lead to improvements, while the full potential of the approach remains to be seen.

\FloatBarrier

\section{Conclusion}
\label{sec:conclusion}
In this work, a Markov decision process (MDP) to model the generation of spatially distributed, entangled quantum states via a memory-based quantum repeater was developed. In principle, the model can describe quantum repeaters with arbitrary numbers of segments and include arbitrary qubit Pauli and erasure channels as error sources. Moreover, entanglement swapping and discarding of quantum states are actions available to operate the quantum repeater.

Based on this MDP, a simulation was implemented. Among the above-mentioned general error channels, the particular error sources included in the simulation are the random, time-dependent dephasing of quantum states in the memories and distance-dependent photon losses in the optical quantum channels (as well as constant losses or inefficiencies at the repeater stations and interfaces). The entanglement swapping is assumed to be error-free and deterministic, and all schemes in this paper "swap as soon as possible" (which is the optimal strategy in the presence of memory dephasing). The simulation was used to analyze the secret key rate of the BB84 quantum key distribution (QKD) protocol via four-segment quantum repeaters in a broad parameter space of the segment length and the quantum memories' coherence time. Moreover, the simulation was used to analyze the behavior of the secret key rate under the variation of a controlled limited storage time of the quantum states - the so-called memory cut-off - in the same parameter space. It was found that there exists a parameter regime in the limit where the memory coherence times are so large or small in relation to the segment length that the ideal control is to discard no or all intermediate quantum states, respectively.

Furthermore, a deep reinforcement learning (DRL) algorithm was implemented to examine the possibilities of finding sophisticated solutions for the control of the quantum memories improving the secret key rate of quantum repeaters. The algorithm of choice was a proximal policy optimization (PPO) \cite{schulman2017proximal}, which was applied to the aforementioned simulated quantum repeaters. First, the results of the limit parameter regime were successfully reproduced. More specifically, and most importantly, the algorithm found policies outperforming the benchmarks provided by the previously employed, "naive" simulation (i.e., a standard simulation without the help of a learning agent), serving as a proof of concept that DRL can indeed offer a valid approach to optimize the memory storage times. This proof of concept is a first step in laying the groundwork to develop and apply DRL algorithms to realistic and practical long-distance quantum key distribution.

The DRL algorithm introduced and employed here, though adapted to the special problem of computing secret key rates in quantum repeaters subject to memory dephasing, is not yet an optimal algorithm. First, it suffers from rather slow convergence properties due to the sparse entanglement distribution per time step for realistic parameters and the lack of an additive reward model. Hence, a next step would be to increase computational power to improve the efficiency with which policies can be found. A further suggestion is to modify the experience loop of the algorithm in a way that flattens the distribution of environment states the agent interacts with to improve the balance of the gradient steps for the different states.

Examining the achieved policies, it is clear that these are not perfect. This suggests that there still is further potential in the approach, even for the examined repeater constellations. In the future, one could expand the simulation to more realistic scenarios, for example, by including more error sources and classical communication, therefore finding solutions for practical settings. In conclusion, the full potential of optimizing policies for quantum repeaters using DRL remains to be seen.

Long-distance key distribution based on quantum repeaters and DRL are both rapidly moving and developing fields, with their full future impact being unforeseeable. This work contributes new insights by combining the two fields and showing viability, as well as problems and limitations of the approach.

\begin{acknowledgments}
We thank the BMBF in Germany for support via PhotonQ, Q.Link.X/QR.X and the BMBF/EU for support via QuantERA/ShoQC. We also thank Frank Schmidt and Evgeny Shchukin for helpful discussions.
\end{acknowledgments}

\appendix

\section{Markov decision process modeling a quantum repeater}
\label{sec:markov-decision-process-modeling-a-quantum-repeater}
Here we will formulate the model by explaining each element of the tuple $(S,A,P_a,R_a)$.

\subsection{States $S$}
Any state of the Markov model is fully described by the spatially separated bipartite quantum states stored in the quantum memories. Thus, in a general treatment, the full density matrices of all quantum states would be necessary to determine the state of the MDP. By restricting the errors to be Pauli channels as defined in Eq. (\ref{eq:pauli-channel}), it is possible to define the states in a much more convenient form. In this restricted scenario, it follows directly from the fact that Pauli channels and entanglement swapping commute, that treating the precise development of all quantum states is equivalent to treating a noiseless quantum repeater and applying any contributing errors to the final quantum state. That means it is sufficient to keep track of any error occurring and propagating the counts additive through the swapping. Therefore, all possible states of the MDP can be encoded in a triangular matrix $S$, with rows and columns corresponding to the repeater nodes. The entries $s_{ij}$ are either the symbol $\xi$, indicating that the two repeater nodes $i$ and $j$ do not share an entangled state or a vector with an integer component for each error, storing its accumulated count on the bipartite quantum state.

Generally, the number of possible states is infinite. In realistic scenarios, the memory strategy restricts the transitions between the states in a way that only a finite amount of states remains accessible with non-vanishing probability.

\subsection{Actions $A$}
The set of possible actions $A$ can be separated into two independent sets of actions, $A = A_s \bigotimes A_d$.
\begin{itemize}
\item The first set corresponds to entanglement swapping. It is encoded in a vector $A_s$ with a boolean component for each node. If a component is true, the entanglement swapping operation is performed at that node. Otherwise, no action will be performed at that node.
\item The second set corresponds to the discarding of states. It is encoded in a triangular matrix $A_d$ with a boolean entry for each pair of two different nodes. If an entry is true, the bipartite quantum state of this pair is discarded. Otherwise, no operation will be performed for this quantum state.
\end{itemize}
It is important to note that in the treatment of the MDP, these actions are instantaneous. The reason for this is that the MDP only models the development of the quantum states, which is independent of classical communication. That does include entanglement swapping. Even though entanglement swapping requires classical communication, the time when classical communication is performed is irrelevant for the development of the quantum states and can thus be excluded from the treatment.

\subsection{Transitional properties $P_a$ and $R_a$}
To find a complete, analytic expression for $P_a$ and $R_a$ is an infeasible task. Fortunately, this is not necessary to simulate the process. It is sufficient to find mathematical operations realizing the simulated processes on the states. The time step of the process is defined by a round of classical communication, which is the time it takes for a node to send classical information to an adjacent node. This time step is the time a quantum state will be stored when initially distributed in one segment before the corresponding nodes communicated the successful distribution. In the following, it will be described how each process that takes place in one time step is simulated.
\begin{enumerate}
\item Initial entanglement generation in each segment:\\Within one time step in each segment, an attempt to distribute initial entanglement is performed. This is simulated via sampling in each segment respective to an entanglement generation rate $p$.
\item Natural, uncontrolled development of the quantum states:\\The count of any error that occurs on stored states is increased by one in the tuples in the state matrix $S$. The for this work relevant example is the dephasing channel.
\item Performing the actions:
\begin{enumerate}
\item Entanglement swapping at each station $j$ where $\left( A_s\right) _{j}$ is true :\\For any tuple $\{ i,k \}$, where $s_{i,j} \neq \xi$ and $s_{j,k} \neq \xi$:
\begin{enumerate}
\item $s_{i,k} \leftarrow s_{i,j} + s_{j,k}$, where "$+$" is the element wise addition of the two vectors. This follows from the fact that Pauli channels and entanglement swapping commute with each other.
\item $s_{i,j} \leftarrow \xi$\\$s_{j,k} \leftarrow \xi$
\item Increase the count of any error that occurs due to entanglement swapping by one in the vector $s_{i,k}$.
\end{enumerate}
\item State discarding:\\ For each tuple $\{ i,j \}$ where $\left( A_d\right) _{i,j}$ is true, set $s_{i,j} \leftarrow \xi$.
\end{enumerate}
\item Return the immediate reward $r=F\left( s_{0,n} \right)$ and set $s_{0,n} \leftarrow \xi$. Where $F\left( s \right)$ is the fidelity between the quantum state determined by the tuple $s$ and the quantum state, a noiseless repeater would have distributed. This could be kept more general, by instead returning the vector $s_{0,n}$ and thus the entire quantum state, but this was not necessary for this work.
\end{enumerate}

\subsection{Example step of the MDP}
\begin{figure*}
\makebox[\textwidth][c]{
	\begin{subfigure}[b]{0.4\textwidth}
		\caption{}
		\includegraphics[width=\textwidth]{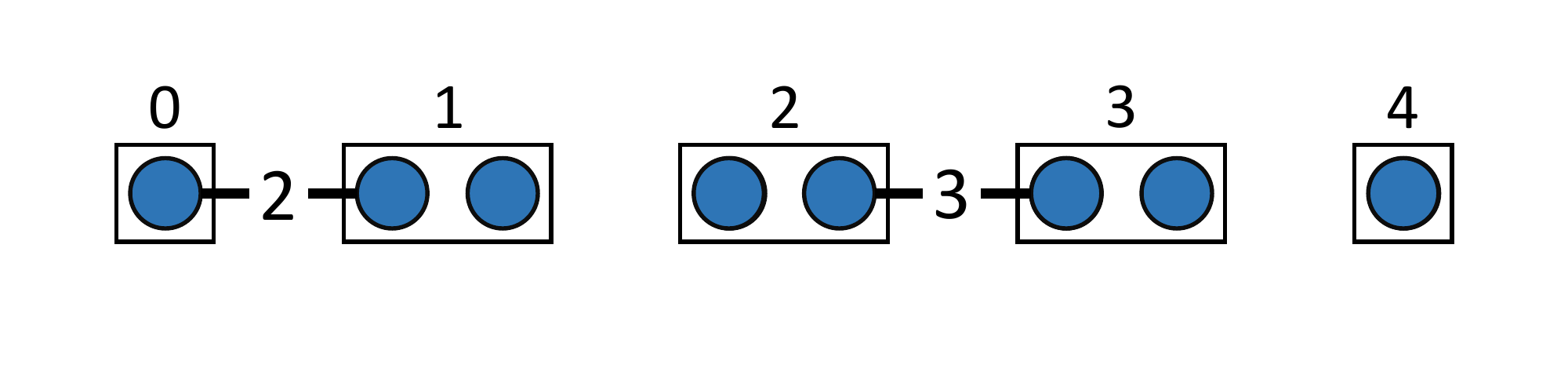}
		\includegraphics[width=\textwidth]{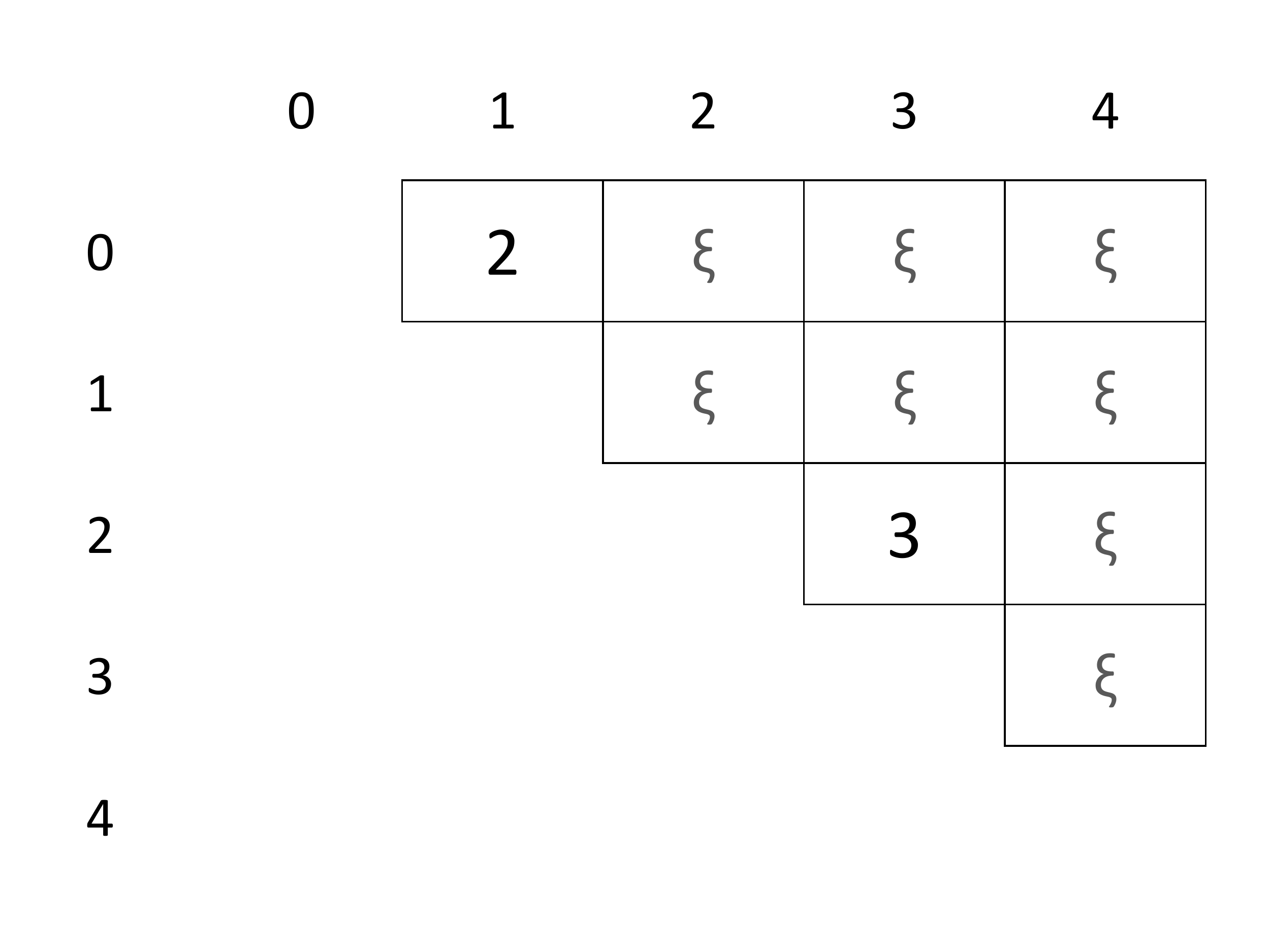}	
		\label{fig:illustration-mdp-a}
	\end{subfigure}
	\begin{subfigure}[b]{0.4\textwidth}
		\caption{}
		\includegraphics[width=\textwidth]{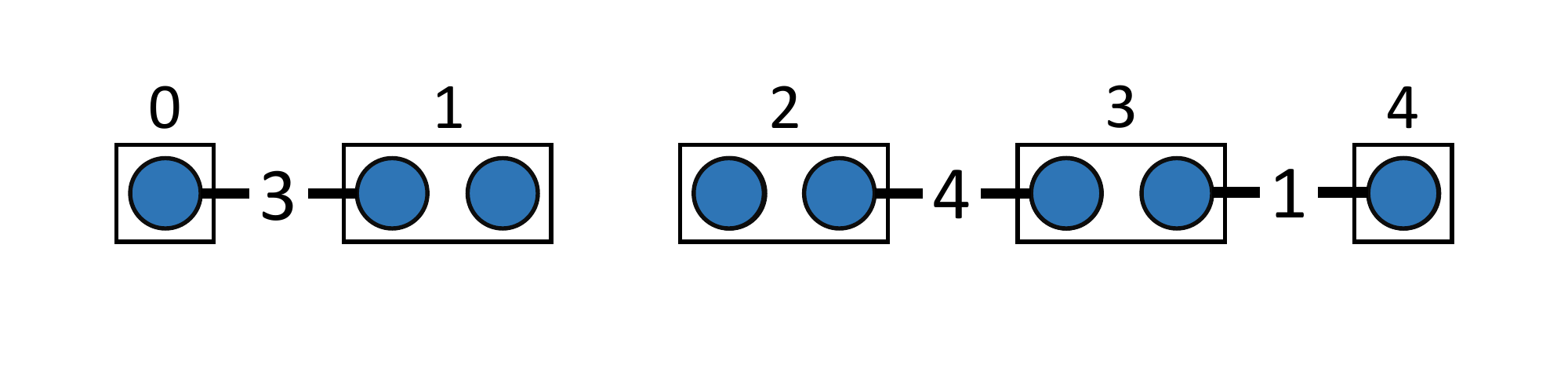}
		\includegraphics[width=\textwidth]{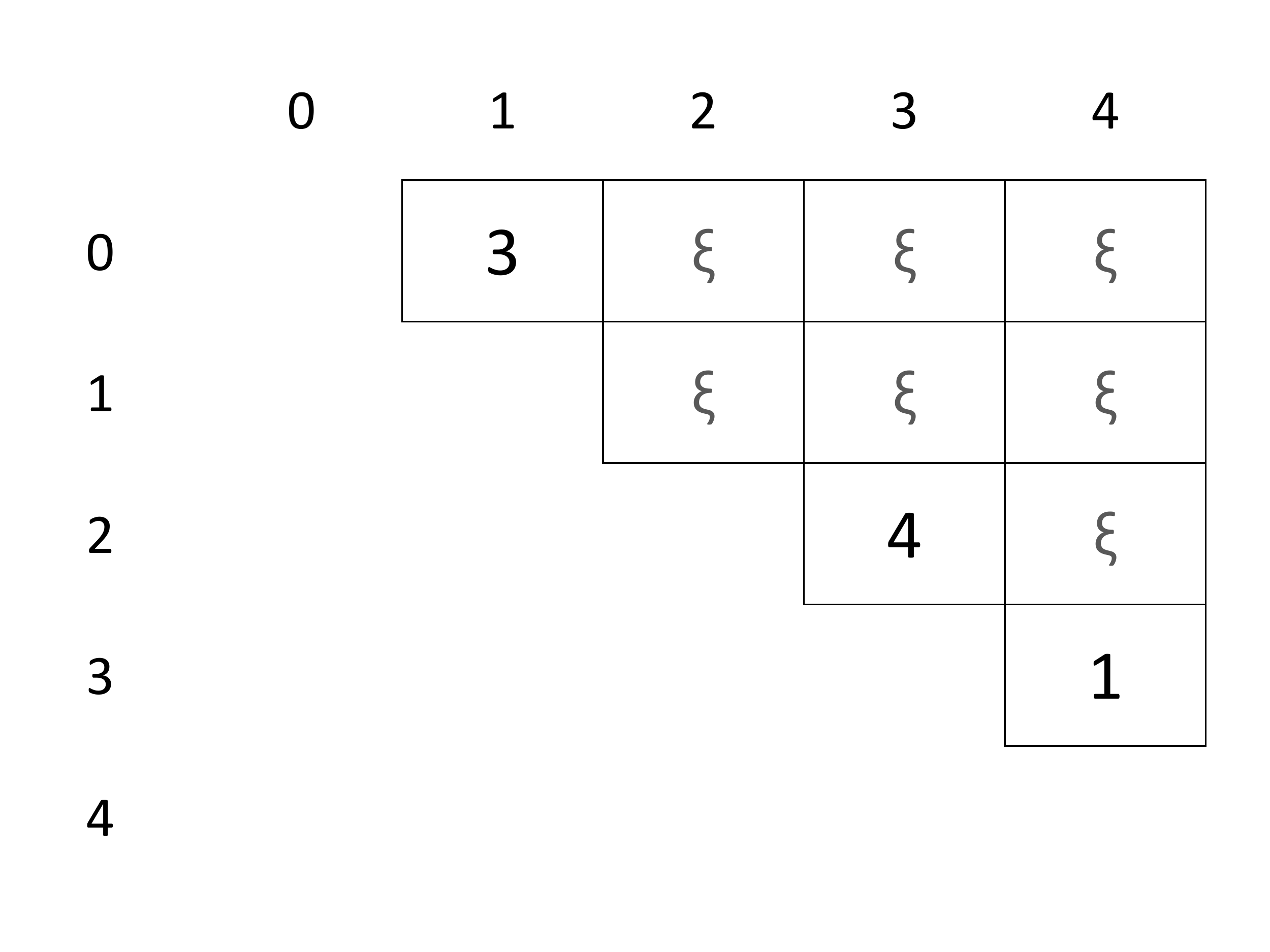}
		\label{fig:illustration-mdp-b}
	\end{subfigure}	
}
\makebox[\textwidth][c]{
	\begin{subfigure}[b]{0.4\textwidth}
		\caption{}
		\includegraphics[width=\textwidth]{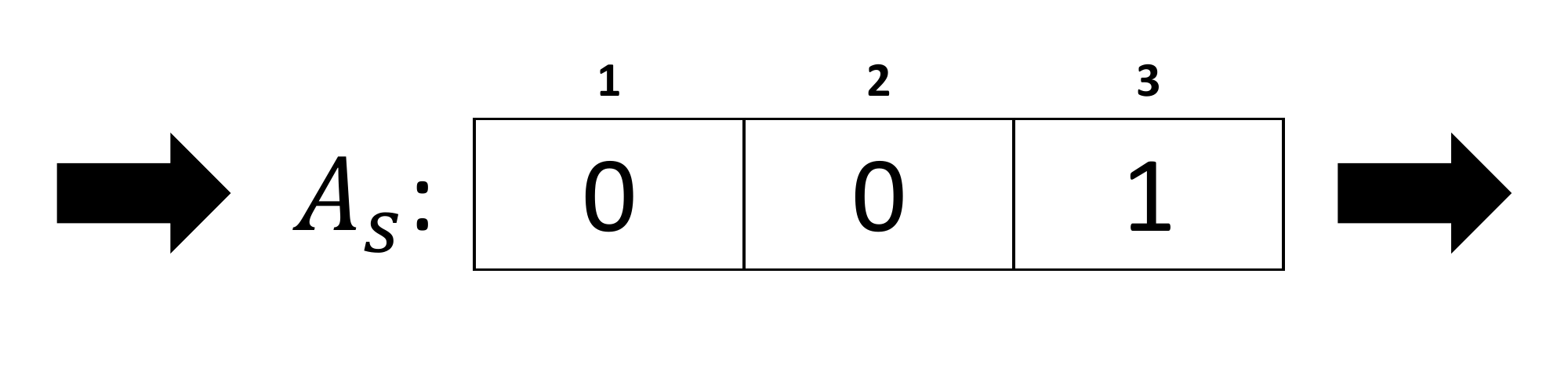}
		\includegraphics[width=\textwidth]{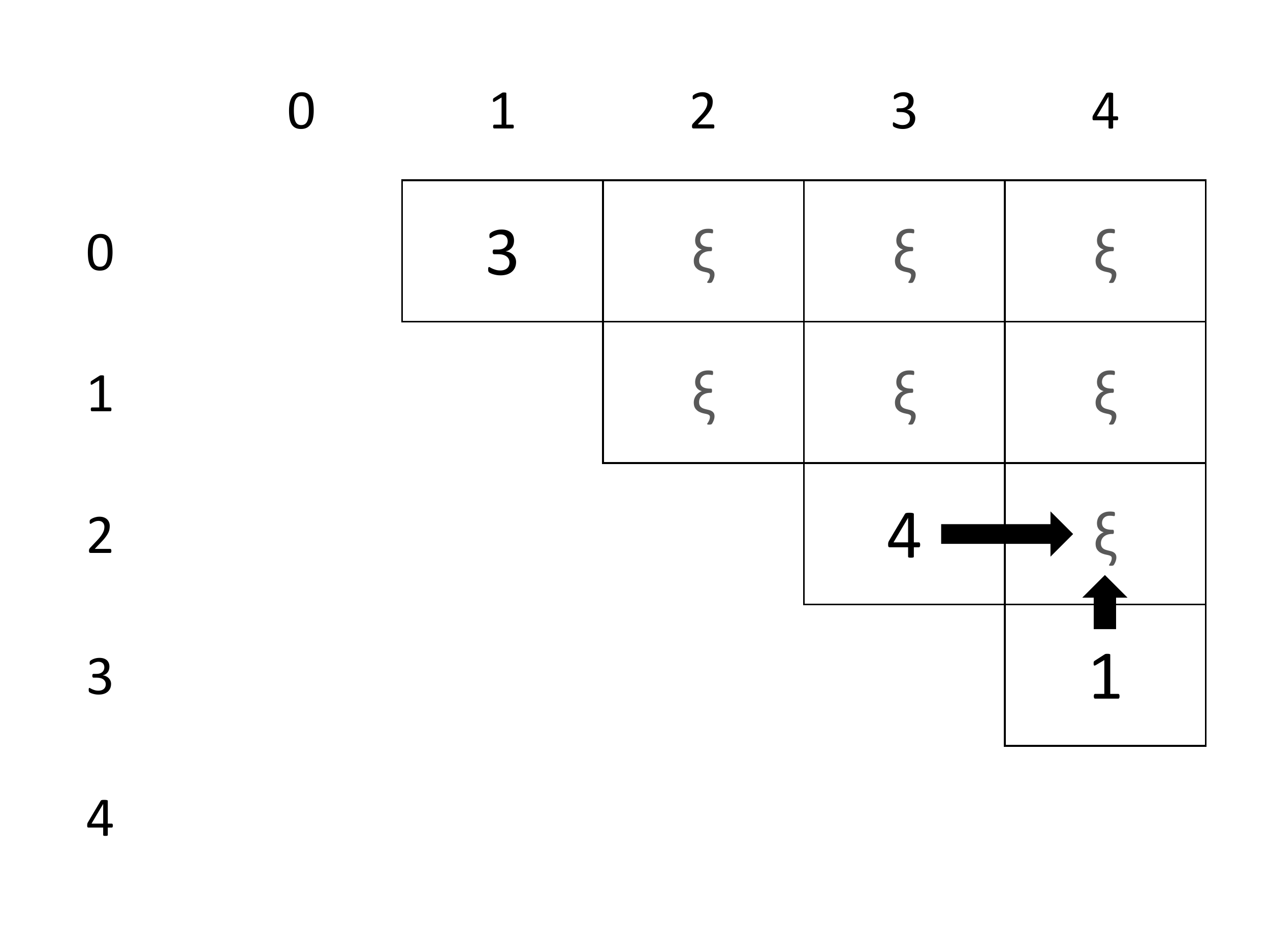}
		\label{fig:illustration-mdp-c}
	\end{subfigure}
	\begin{subfigure}[b]{0.4\textwidth}
		\caption{}
		\includegraphics[width=\textwidth]{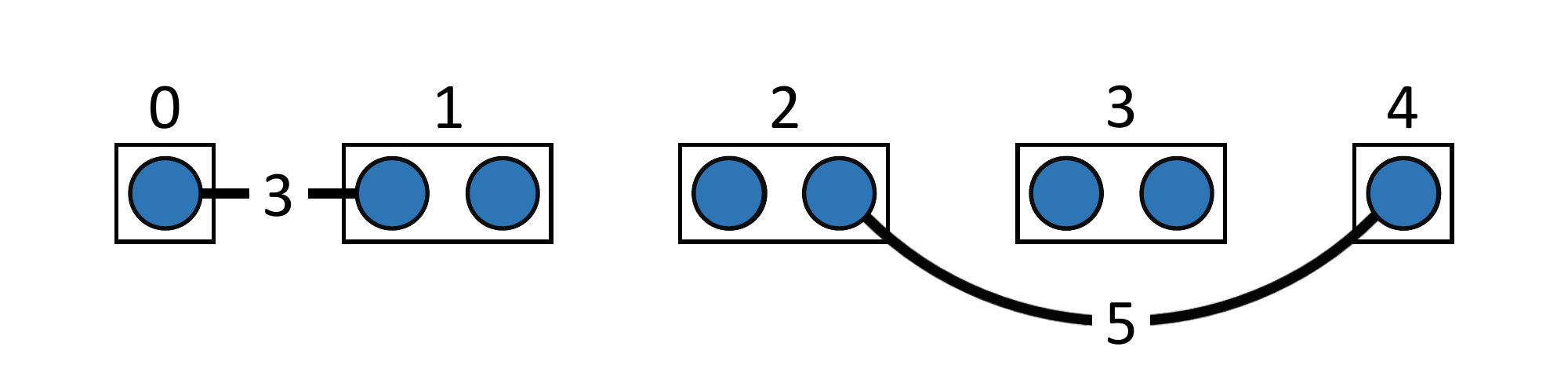}
		\includegraphics[width=\textwidth]{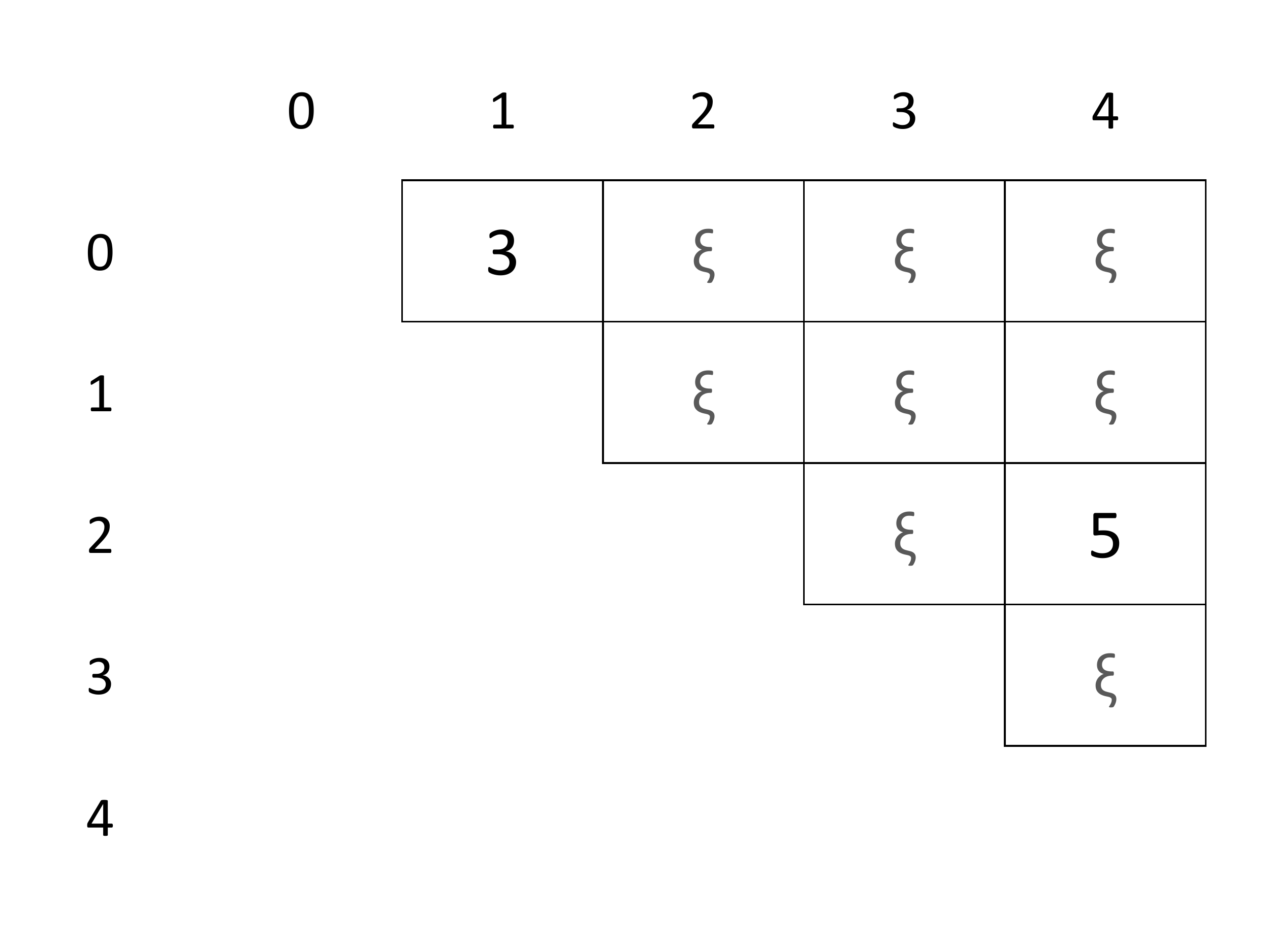}
		\label{fig:illustration-mdp-d}
	\end{subfigure}	
}
\makebox[\textwidth][c]{
	\begin{subfigure}[b]{0.4\textwidth}
		\caption{}
		\includegraphics[width=\textwidth]{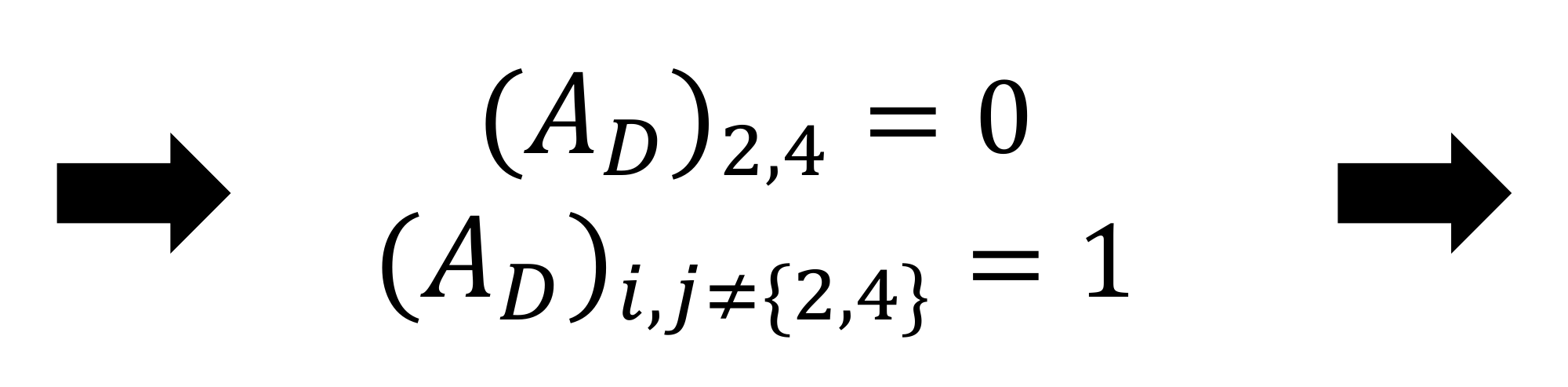}
		\includegraphics[width=\textwidth]{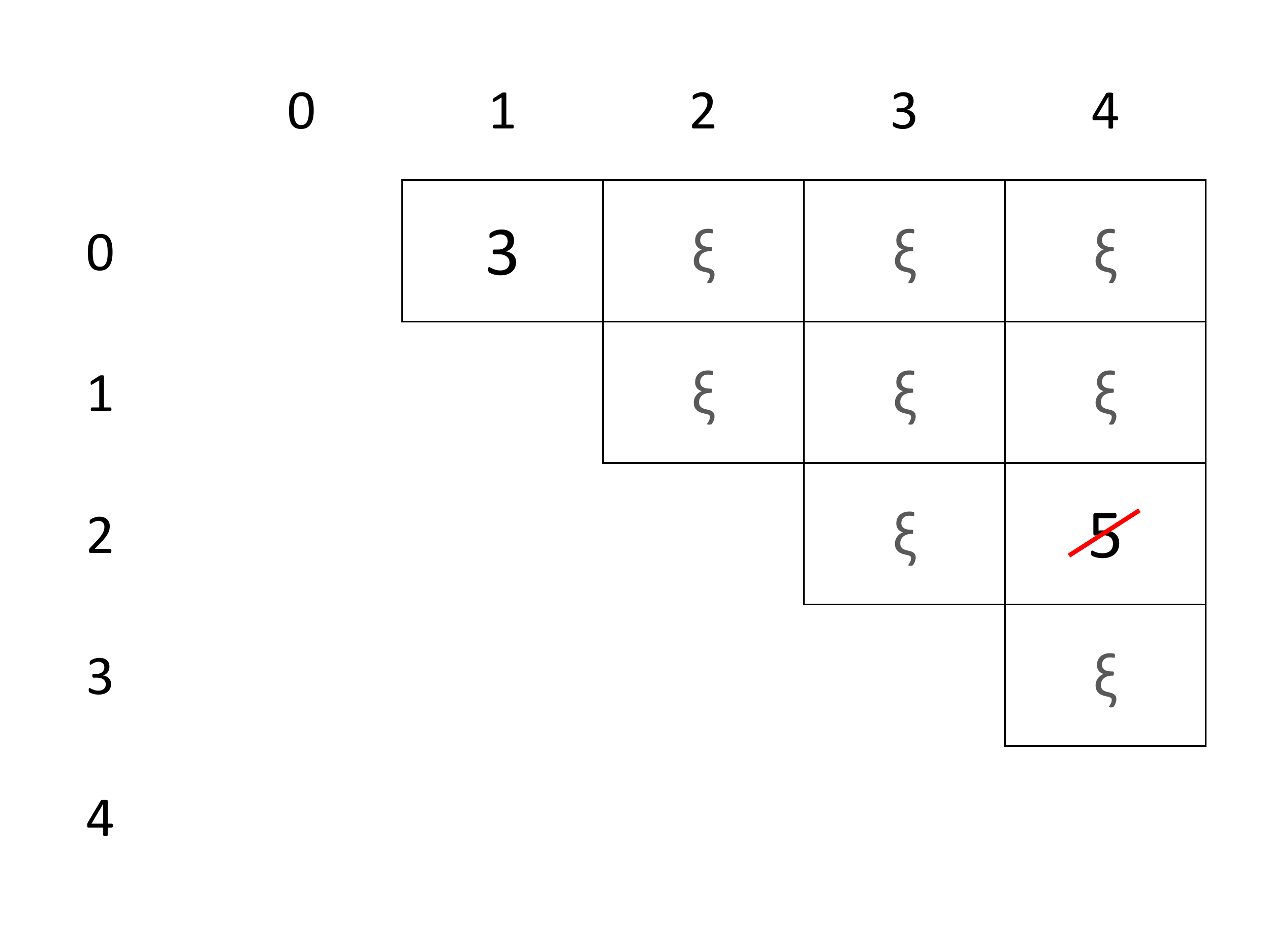}
		\label{fig:illustration-mdp-e}
	\end{subfigure}
	\begin{subfigure}[b]{0.4\textwidth}
		\caption{}
		\includegraphics[width=\textwidth]{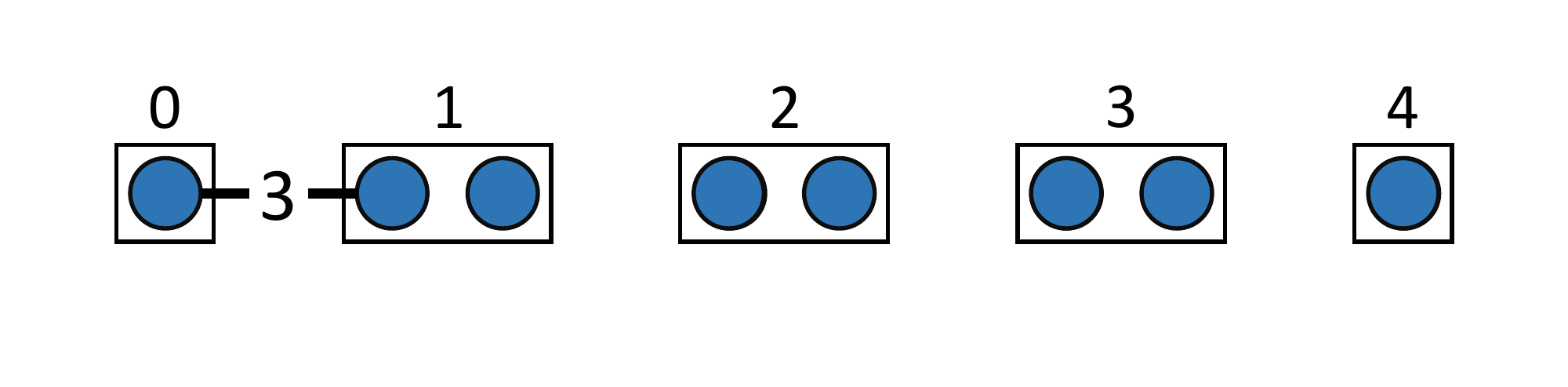}
		\includegraphics[width=\textwidth]{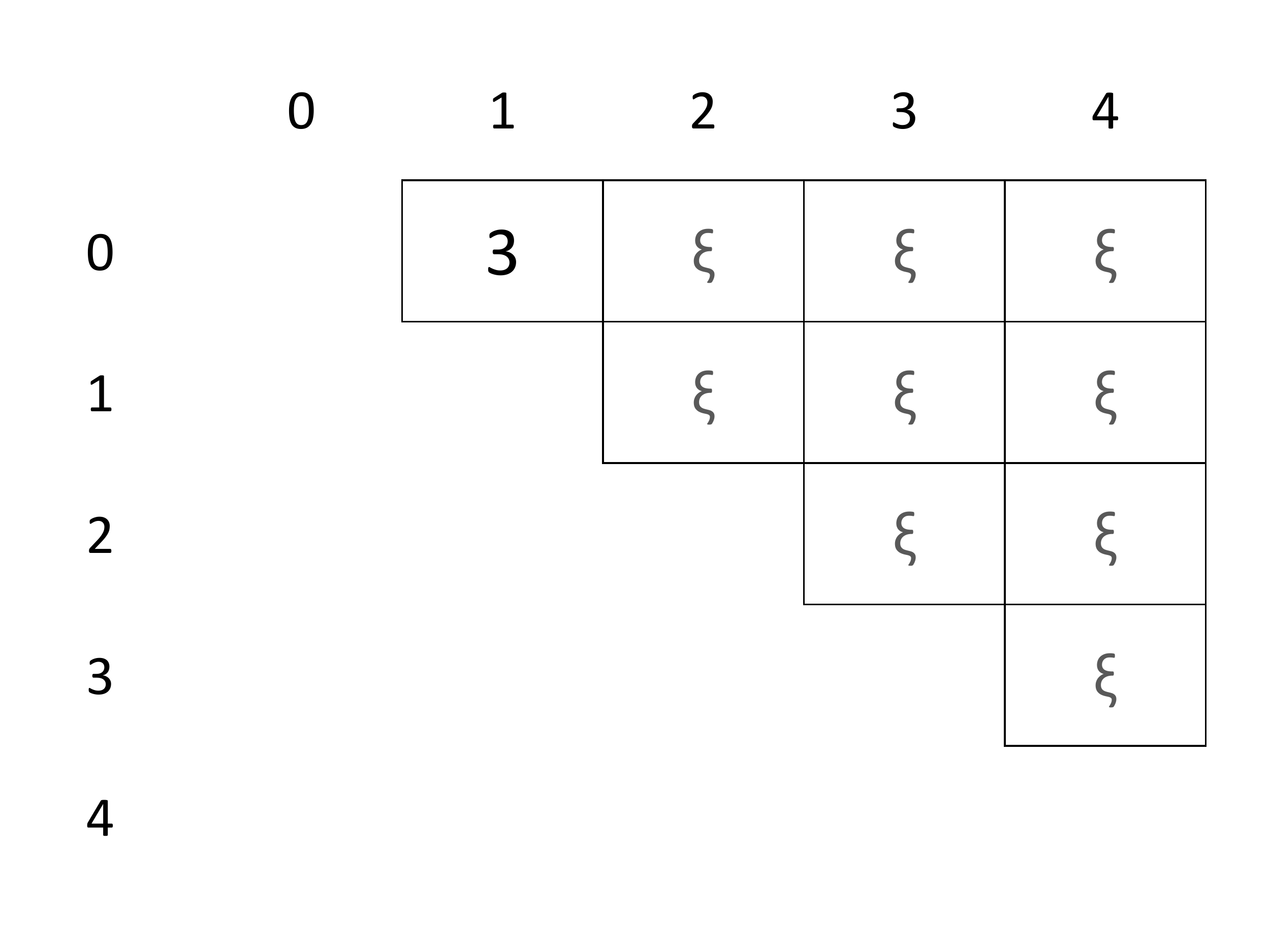}
		\label{fig:illustration-mdp-f}
	\end{subfigure}
}
	\caption{\label{fig:illustration-mdp}Illustration of one example step of the Markov decision process.}
\end{figure*}

In Fig. \ref{fig:illustration-mdp} one example step of the MDP is illustrated. In this example, the only error is the dephasing of the quantum memories, which accumulates on a state for every time step it is stored. Therefore the entries $s_{j,k}$ are the cumulated storage time of the quantum state stored in the nodes $j$ and $k$.

The state of the MDP prior to the step is shown in Fig. \ref{fig:illustration-mdp}(\subref{fig:illustration-mdp-a}). The state after the uncontrolled development, which consists of generating initial entanglement and accumulating the storage time is shown in Fig. \ref{fig:illustration-mdp}(\subref{fig:illustration-mdp-b}). The controlled swapping operation is depicted in Fig. \ref{fig:illustration-mdp}(\subref{fig:illustration-mdp-c}) with the resulting state in Fig. \ref{fig:illustration-mdp}(\subref{fig:illustration-mdp-d}). The discarding action is displayed in Fig. \ref{fig:illustration-mdp}(\subref{fig:illustration-mdp-e}) with the resulting state in Fig. \ref{fig:illustration-mdp}(\subref{fig:illustration-mdp-f}).

\section{Secret key rate of entanglement-based BB84 in the presence of dephasing}
\label{sec:calculation-of-the-secret-key-rates}

In the following, we will calculate the secret key rate for a BB84 protocol where the quantum states are subject to a dephasing channel.

Without loss of generality, we will choose the initially distributed, undisturbed entangled states as $\ket{\Phi^+} = \frac{1}{\sqrt{2}}\left( \ket{00}+\ket{11} \right)$. The form of a degraded state follows directly from Eq. (\ref{eq:dephasing}) and reads
\begin{equation}
	\rho = (1 - \nu) \ket{\Phi^+} \bra{\Phi^+} + \nu \ket{\Phi^-} \bra{\Phi^-},
\end{equation}
where 
\begin{equation}
	0 < \nu < \frac{1}{2}.
\end{equation}

The bit error rates in the $X$ and $Z$ bases are defined as
\begin{equation}
 e_Z = \mathds{E} \left[ 1 - \bra{00} \rho \ket{00} + \bra{11} \rho \ket{11} \right]
\end{equation}
and
\begin{equation}
 e_X = \mathds{E} \left[ 1 - \bra{++} \rho \ket{++} + \bra{--} \rho \ket{--} \right].
\end{equation}

A straightforward calculation yields
\begin{equation}
	e_Z = 0
\end{equation}
and
\begin{equation}
	e_X = \mathds{E} \left[ \nu \right].
\end{equation}

Inserting the bit error rates into Eq. (\ref{eq:BB84}) and Eq. (\ref{eq:skfraction}) gives
\begin{equation}
	R_{\text{BB84}} = Y \cdot \left( 1 - h\left( \mathds{E} \left[ \nu  \right] \right) \right).
	\label{eq:RBB84}
\end{equation}

Alternatively, by inserting the exponential dephasing channel defined in Eq. (\ref{eq:dephasing}), Eq. (\ref{eq:RBB84}) takes the form
\begin{equation}
	R_{\text{BB84}} = Y \cdot \left( 1 - h\left( \frac{1}{2} \left( 1 - \mathds{E} \left[ e^{- \frac{t}{\tau_c}} \right] \right) \right) \right),
\end{equation}
where $\tau_c$ is the coherence time of the quantum memory for a bipartite quantum state and $t$ the storage time of the quantum state.\footnote{In practical finite-size applications one can use confidence intervals on the estimation of the expectation value to determine an error rate which is larger than the true expectation value with approximately guaranteed probability.}

\section{Discounting rewards}
\label{sec:A:discounting}
\subsubsection{Finite and infinite horizons and the relevance of discounting rewards}
The in Sec. \ref{sec:deep-reinforcement-learning:alg} introduced discounted reward function of Eq. (\ref{eq:Rt}) is a fundamental objective function in RL:
\begin{equation}
	R_t = \sum^{T-t}_{l=0} \gamma^l r_{t+l}.
\end{equation}
A discounting parameter $\nu$ smaller one introduces the discounting into the accumulated reward. This servers to intuitive purposes. First, it ensures that the objective function remains finite for "infinite-horizon" trajectories (T $\rightarrow \infty$). Secondly, and maybe most importantly for most practical applications, it makes the agent weight sooner rewards more than rewards in the farther future. This improves the evaluation of an action as it is bias towards its short-term consequences rather than on rewards in the distant future, which is increasingly independent from the action at the time $t$. In this way, discounting reduces noise in the objective function.

As was stated before, the non-additive rewards in this work are not suited for discounting in this way. Even more detrimental , non-negative rewards can actually decrease the secret key rate of a trajectory. The lack of discounting necessitates the use of a finite-horizon objective function, which makes good gradient estimation difficult. This can be explained by understanding that, when the trajectory is too long, the objective function becomes very noisy, since for earlier actions, the long trajectory introduces rewards that are mostly independent from early actions. On the other hand, if the trajectory is too short, delayed rewards of an action might not be obtained, cause the trajectory ended, causing a wrong evaluation of the action.

In consequence, good convergence is expected to be significantly harder to achieved compared to a problem setting where discounting can be used.

\subsubsection{A proposal for generalized discounting}
In this work an attempt was made to propose discounted objective function of an arbitrary (not necessarily additive) reward function $\Psi(\tau)$ via:
\begin{equation}
	\Phi_t = \sum_{l=0}^{T-t} \gamma^l \Psi(\tau_t).
\end{equation}
Applied to our optimization this reads,
\begin{equation}
	\Phi_t = \sum_{l=0}^{T-t} \gamma^l R_\text{BB84}(\tau_t).
\end{equation}
This seemed like a reasonable approach intuitively but yielded no useful results during the minimal testing we did. Therefore, this method was not used further. Even though, unfortunately, this could not be achieved here, with sufficient hyperparameter tuning this method could possibly improve the convergence properties of the optimization significantly, as of the reasons discussed above.

\section{Trajectory lengths of the simulations}
\label{sec:A:sim-trajectory-lengths} 

In this appendix the lengths of the trajectories used in the simulations and the experience collection of the agents is given in Tab. \ref{tab:hyperparameter-sim}.

\begin{table}[h]
\begin{tabular}{llccc}
	\toprule
	$L_0$ (km) & $\tau_c$ (ms) & $T$, no cut-off & $T$, cut-off & $T$, agent \\
	\midrule
	$20$ & $0.1$ & $10^4$ & $10^4$ & $10^5$ \\
	$20$ & $1$ & $10^4$ & $10^4$ & $10^5$ \\
	$20$ & $10$ & $10^4$ & $10^4$ & $10^5$ \\
	$20$ & $100$ & $10^4$ & $10^4$ & \\
	$35$ & $0.1$ & $10^4$ & $10^4$ & \\
	$35$ & $1$ & $10^4$ & $10^4$ & $10^5$ \\
	$35$ & $10$ & $10^4$ & $10^4$ & $10^5$ \\
	$35$ & $100$ & $10^4$ & $10^4$ & \\
	$50$ & $0.1$ & $10^6$ & $10^6$ & \\
	$50$ & $1$ & $10^5$ & $10^4$ & $10^5$ \\
	$50$ & $10$ & $10^5$ & $10^4$ & $10^5$\\
	$50$ & $100$ & $10^4$ & $10^4$ & $10^5$\\
	$50.6569$ & $1.45271$ & $10^5$ & $10^5$ & $10^5$ \\
	$70$ & $0.1$ & $10^5$ & $10^6$ & \\
	$70$ & $1$ & $10^5$ & $10^5$ & \\
	$70$ & $10$ & $10^5$ & $10^5$ & \\
	$70$ & $100$ & $10^5$ & $10^5$ & \\
	\bottomrule
	\sbline
\end{tabular}
	\caption{\label{tab:hyperparameter-sim}Length of the trajectories of the simulations.}
\end{table}

\FloatBarrier

\section{Hyperparamters of the DRL runs}
\label{sec:A:drl-hyperparameter} 
The DRL program was run with four processes on a desktop computer with a four-core cpu.\footnote{Intel(R) Core(TM) i7-3770 CPU @ 3.40GHz} An epoch with four trajectories with each 8000 simulated time steps took about 13 seconds on average. Therefore, a thousand epochs take around 3.6 hours. In the operated hyperparameter space, the computation time was roughly linear with the number of simulated time steps within the epoch. This serves as an orientation for the required computation time of each run.

The architecture of the neural networks representing the policy and the value function was chosen identical for every learning run as it seemed sufficient in complexity and degrees of freedom for the task, and smaller architectures did not improve computation time significantly, empirically. The architecture consisted of 2 hidden layers with 32 neurons each. The activation function of the hidden layers was chosen to be the tangens hyperbolicus. The neural network representing the policy applies a sigmoid as the activation on the output layer so that the final output can be interpreted as probabilities.

Abbreviations for the hyperparameters:

\begin{itemize}
\item L: Length of the simulated trajectories.
\item N: Number of simulated trajectories per epoch.
\item $\alpha_{\pi}$: Learning rate for the policy.
\item $\alpha_{V}$: Learning rate for the value function.
\item $\epsilon_{\text{clip}}$: Epsilon of the clipped objective of PPO in Eq. (\ref{eq:PPO-alg}).
\item $n_{\pi}$: Number of gradient steps in the policy update in one epoch.
\item $n_{V}$: Number of gradient steps in the value function update in one epoch.
\item KL: Maximum KL-divergence which stops the gradient update steps of the policy early if exceeded.
\item $\epsilon_{\text{Adam}}$: The parameter in the Adam optimization to improve numerical stability.
\end{itemize}

\begin{table*}[tb]
\begin{tabular}{lllccccccccc}
	\toprule
	figure & $L_0$ (km) & $\tau_c$ (ms) & L & N & $\alpha_{\pi}$ & $\alpha_{V}$ & $\epsilon_{\text{clip}}$ & $n_{\pi}$ & $n_{V}$ & KL & $\epsilon_{\text{Adam}}$\\
	\midrule
	\ref{fig:4-20-100-drl} & $20$ & $0.1$ & $800$ & $4$ & $1 \cdot 10^{-4}$ & $1 \cdot 10^{-3}$ & $0.2$ & $120$ & $80$ & $0.015$ & $0.01$\\ \addlinespace
	\ref{fig:4-20-1000-drl} & $20$ & $1$ & $8000$ & $4$ & $4 \cdot 10^{-4}$ & $1 \cdot 10^{-3}$ & $0.2$ & $120$ & $80$ & $0.015$ & $0.01$\\ \addlinespace
	\ref{fig:4-20-10000-drl} & $20$ & $10$ & $8000$ & $4$ & $1 \cdot 10^{-3}$ & $1 \cdot 10^{-3}$ & $0.2$ & $120$ & $80$ & $0.015$ & $0.01$\\ \addlinespace
	\ref{fig:4-35-1000-drl} & $35$ & $1$ & $8000$ & $4$ & $1 \cdot 10^{-3}$ & $1 \cdot 10^{-3}$ & $0.2$ & $120$ & $80$ & $0.015$ & $0.01$\\ \addlinespace
	\ref{fig:4-35-10000-drl} & $35$ & $10$ & $8000$ & $4$ & $4 \cdot 10^{-4}$ & $1 \cdot 10^{-3}$ & $0.2$ & $120$ & $80$ & $0.015$ & $0.01$\\ \addlinespace
	\ref{fig:4-50-1000-drl} & $50$ & $1$ & $8000$ & $4$ & $8 \cdot 10^{-4}$ & $1 \cdot 10^{-3}$ & $0.2$ & $1000$ & $80$ & $0.015$ & $0.1$\\ \addlinespace
	\ref{fig:4-50-10000-drl} & $50$ & $10$ & $8000$ & $4$ & $4 \cdot 10^{-3}$ & $1 \cdot 10^{-3}$ & $0.2$ & $1000$ & $80$ & $0.015$ & $1$\\ \addlinespace
	\ref{fig:4-50-100000-drl} & $50$ & $100$ & $8000$ & $4$ & $4 \cdot 10^{-3}$ & $1 \cdot 10^{-3}$ & $0.2$ & $1000$ & $80$ & $0.015$ & $1$\\ \addlinespace
	\ref{fig:4-50-1452-drl} & $50.6569$ & $1.45271$ & $8000$ & $4$ & $4 \cdot 10^{-4}$ & $1 \cdot 10^{-3}$ & $0.2$ & $120$ & $80$ & $0.015$ & $1 \cdot 10^{-8}$\\
	\bottomrule
	\sbline
\end{tabular}
	\caption{\label{tab:learning-reproduced-hp}Hyperparameters of the learning runs presented in Sec. \ref{sec:drl-results}}
\end{table*}

\FloatBarrier

\bibliographystyle{apsrev4-2} 
\bibliography{Deep-reinforcement-learning-for-key-distribution-based-on-quantum-repeaters-bib-refs} 

\end{document}